\documentclass[sigconf]{acmart}
\AtBeginDocument{%
  \providecommand\BibTeX{{%
     \normalfont B\kern-0.5em{\scshape i\kern-0.25em sb}\kern-0.8em\TeX}}}
     
\usepackage{tikz}
\usepackage{amsmath}
\usepackage{amsthm}
\usepackage{amsfonts}
\usepackage{algorithmic}
\usepackage{algorithm}
\usepackage{booktabs}
\usepackage{graphicx}
\usepackage{makecell}
\usepackage{multirow}
\usepackage{mathrsfs}
\usepackage{xcolor}

\begin{document}
\title{Auto-MAP: A DQN Framework for Exploring \\Distributed Execution Plans for DNN Workloads}

\author{Siyu Wang, Yi Rong, Shiqing Fan, Zhen Zheng,} \author{LanSong Diao, Guoping Long, Jun Yang, Xiaoyong Liu, Wei Lin}

\affiliation{Alibaba Group}
\affiliation{\{siyu.wsy, rongyi.ry, shiqing.fsq, james.zz, \\lansong.dls, guopinglong.lgp, muzhuo.yj, xiaoyong.liu, weilin.lw\}@alibaba-inc.com}

\renewcommand{\shortauthors}{X.et al.}

\begin{abstract}
The last decade has witnessed growth in the computational requirements for training deep neural networks. Current approaches (e.g., data/model parallelism, pipeline parallelism) parallelize training tasks onto multiple devices. However, these approaches always rely on specific deep learning frameworks and requires elaborate manual design, which make it difficult to maintain and share between different type of models.
In this paper, we propose Auto-MAP, a framework for exploring distributed execution plans for DNN workloads, which can automatically discovering fast parallelization strategies through reinforcement learning on IR level of deep learning models. Efficient exploration remains a major challenge for reinforcement learning. We leverage DQN with task-specific pruning strategies to help efficiently explore the search space including optimized strategies. Our evaluation shows that Auto-MAP can find the optimal solution in two hours, while achieving better throughput on several NLP and convolution models.
\end{abstract}

\keywords{deep learning, data parallelism, pipeline parallelism, model parallelism, DQN algorithm, HLO IR.}
\settopmatter{printacmref=false} 
\renewcommand\footnotetextcopyrightpermission[1]{} 
\pagestyle{plain} 

\makeatletter
\renewcommand\@formatdoi[1]{\ignorespaces}
\makeatother

\maketitle
\section{Introduction}
\label{sec:intro}

Deep learning (DL) models has become increasingly complicated in artificial intelligence (AI) community to acquire better accuracy. 
Training deep models is extremely both time and resources consuming. Distributed training with multiple devices is a irreversible trend for training especially for large models. \cite{AI-and-compute, rajbhandari2019zero}. To harness computing power to achieve better throughput, a critical challenge is how to map diversified workloads to hardware accelerators automatically and efficiently.

Existing solutions like data parallelism, model parallelism and pipeline parallelism make trade-offs between computation, communication, and development efficiency.
Data Parallelism (DP) is workload-neutral to models that could be fit into single device while facing the problem of memory footprint pressure for large models.
Model parallelism (MP) \cite{shoeybi2019megatron, shazeer2018mesh, jia2018exploring, geng2019horizontal, dryden2019channel, lepikhin2020gshard} and pipeline parallelism (PP) \cite{huang2019gpipe, narayanan2019pipedream} are effective way for alleviating the memory issue of large models, which split the model among processes, in vertical and horizontal way respectively. But the experts experiences are required to design a specific strategy to fully utilize hardware under the limited computation resources.

Some previous works\cite{harlap2018pipedream, raffel2019exploring, Jia2018Beyond} oriented to exploring distributed plans which combine dynamic programming and heuristic methods have been proposed as promising approaches for training complex models. 
But these approaches are designed for specific category of parallelism strategy. \cite{harlap2018pipedream} aims to finding the best solutions of PP in a synchronous way and \cite{raffel2019exploring, Jia2018Beyond} tries to search the OPP. This leads to the limited scenarios mainly because the optimal strategies for diversified workloads are very different, and all these planners do not contain solution space of DP, operator partitioning parallelism (OPP) and PP at the same time. The heuristic method in \cite{raffel2019exploring} results in the lack of generalization in NLP models. Another issue is that the planners are coupled with the specific APIs and deep learning frameworksso that they only take effect in a limited usage. Moreover, the coarse granularity exploration on layer- or operator-level loses the potentials for better solutions. Moreover, to integrate the planner to other frameworks is unfeasible in real-word.

Recently, a trend of machine learning oriented approaches to optimize performance of systems has been receiving increasingly attention in AI research community. \cite{mirhoseini2017device} adopts reinforcement learning to learn the proper parallelism strategies, which inspired researchers to use learning approaches to extract features of deep models. However, it only search for simple model parallelism strategy without OPP and PP solution space for the given workload and clusters at the expense of huge time- and resources-consuming, which leads to non-applicable in industry. 

Above all, we conclude the following deficiencies of these approaches: (1) Limited applicable scenarios, which lacks the coverage of convolution, language model (LM), search/recommendation models at the same time.
(2) Limited parallelism scenarios. None of these efforts have achieved the support of model parallelism, data parallelism, and pipeline parallelism on a unified computing layer (e.g, TF graph).
(3) Inevitable code intrusion. The planners only take effect when specific APIs are called. It fails to to shield users from low-level distributed details.

We propose \emph{Auto-MAP}, a unified framework for exploring distributed execution plans, which works on \textit{HLO IR} via DQN method for DNN workloads. 

\emph{Auto-MAP} works on HLO IR instead of operators or layers. \textit{HLO IR} is an intermediate representation which produced by XLA (Accelerated Linear Algebra) from TensorFlow framework, which describes the entire training task with more general and expressive computation \emph{Instruction}s instead of the operation like GraphDef in TensorFlow. Each instruction contains all necessary information for computation. Some extra information such as the corresponding operator name it belongs to is also recorded. There are two reasons for choosing HLO IR as the operational level of \emph{Auto-MAP}. One is that to explore distributed plans on HLO IR can achieve better performance benefits from its finer granularity than operators. The other is XLA can exist independently from TensorFlow and it has the ability to support other front ends like Jax\cite{jax2018github} and Trax\cite{Trax20}, which leads to no invasion to user codes.

\begin{figure}[ht]
    \centering
    \includegraphics[width=0.20\textwidth]{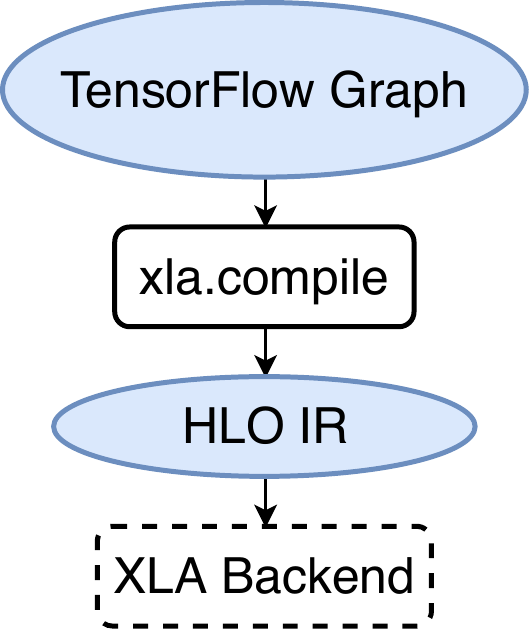}
    \caption{
    Illustration of the high-level design of Tensorflow's XLA compiler.
    }
    \label{fig:tf-xla-hlo}
    \vspace{0.5cm}
\end{figure}

Figure \ref{fig:tf-xla-hlo} gives the high-level design of TF'XLA compiler. As the figure shows, the XLA compiler compiles a TF graph (an ML network in TF) into executable machine code through a sequence of stages.
The TF graph is first transformed into HLO IR by a front-end (e.g., the $xla.compile$ API\cite{xla19}).
Optimizations, such as operator fusion and common-subexpression elimination \cite{muchnick1997advanced} are performed on HLO before the graph is transformed into a lower-level representation for a target hardware architecture.

Deep Q network (DQN) is a reinforcement learning (RL) is the approach to teach machines to interact with the environments and receive rewards for performing the right actions until they successfully meet their goals. It is adopted in \emph{Auto-MAP} to learn the features of deep models and provide workload-neutral distributed plans on given computation resources. It should be noted that the solution space is still huge even with DQN. Therefore, some heuristic pruning methods is also integrated in our approach. As far as we know, there is no previous work focusing on exploring strategies including the three category of parallelism  simultaneously mentioned above with DQN.

\begin{figure}[ht]
    \centering
    \includegraphics[width=0.48\textwidth]{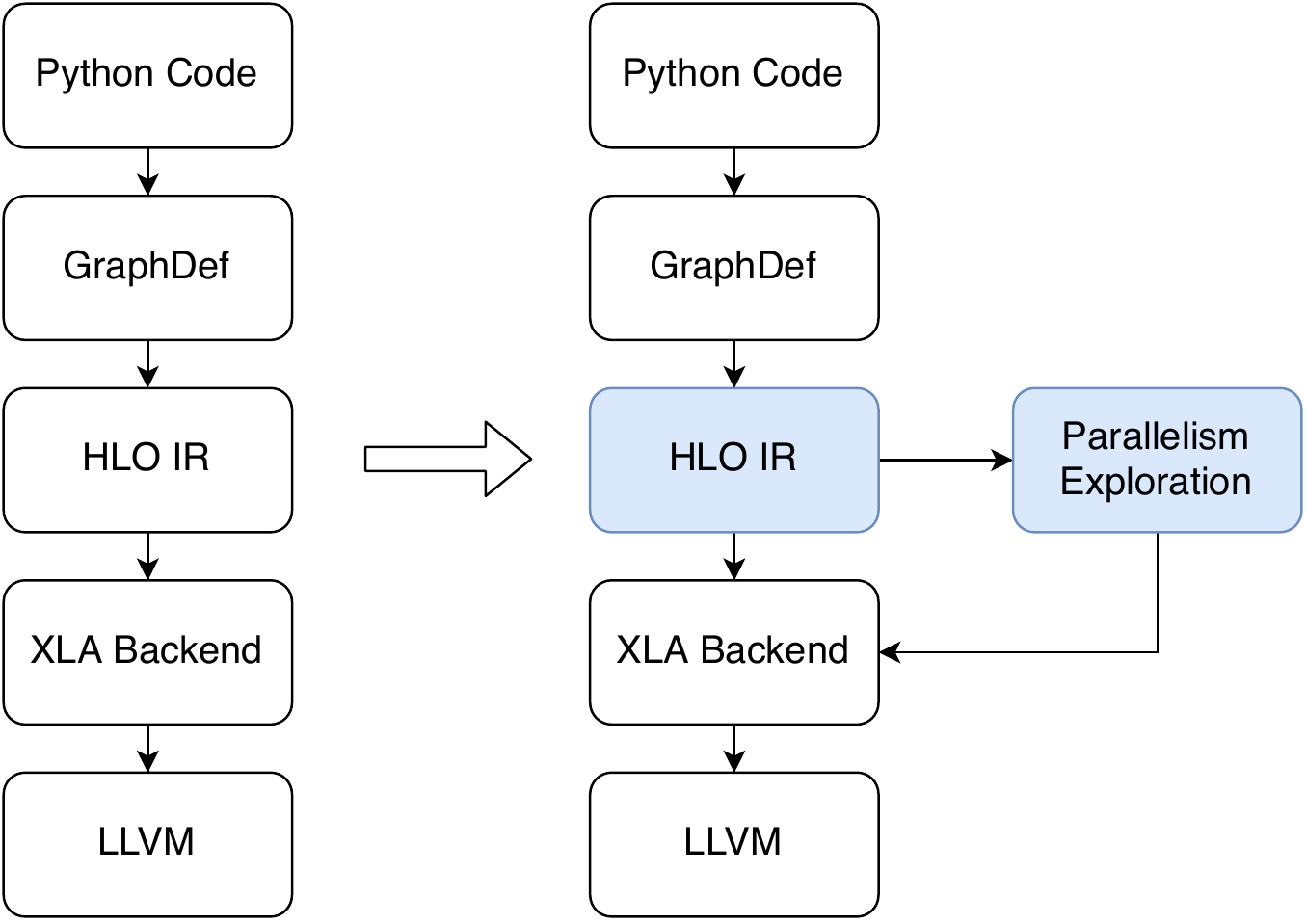}
    \caption{
    Illustration of our approach over TF XLA compiler's work.
    }
    \label{fig:hlo-parallelism}
\end{figure}

As shown in figure \ref{fig:hlo-parallelism}, Auto-MAP performs distributed plans exploration at $HLO$ layer. Compared with previous approach, this has the following advantages: (1) Free user code intrusion. The user only needs to provide a single-device model and the distributed details generated by our Auto-MAP framework are absolutely shielded. (1) Rich and unified parallelism and application scenarios. Unify DP/MP/PP for CNN/LM/Recommendation models. (3) Diverse programming abstractions over HLO IR. Popular AI frameworks such Tensorflow, PyTorch\cite{paszke2017automatic}, Flax\cite{jax2018github}/Trax\cite{Trax20} can all map to $HLO$ layer. In this work, we leverage DQN algorithm \cite{mnih2013playing} to automatically explore the search space of operator partitioning parallelism, auto data parallelism and pipeline parallelism over $HLO$ with device and network interconnect topology specified.

In this paper, we focus on solving the two main challenges of distributing diverse and complex models to distributed heterogeneous hardware platforms: leverage DQN algorithm to build a search space including optimized strategies over $HLO\ IR$, and leverage task-specific pruning method for more efficiently exploration of search space.

To summarize, our contributions are:
\begin{enumerate}
  \item We propose a unified framework named Auto-MAP for three typical parallelism strategies (i.e., operation partitioning, auto data parallel and pipeline) and two typical model types (i.e., CNN and language models);
  \item We leverage DQN with task-specific pruning strategies to help efficiently explore the search space including optimized strategies;
  \item We fully simplifies the burden of users in the selection and implementation of distributed execution plans. With our framework, users only need to provide a single-card graph, and our framework automatically explores the distributed execution plans that is compatible with the hardware computing power/interconnection topology;
  \item We show that our framework can find the optimal solution in a limited time compared to enumeration.
\end{enumerate}

\section{Problem Formulation and Preliminaries}

\textbf{Data and model parallelism} have been widely used by existing deep learning frameworks to distribute the models across devices. Data parallelism is parallelization across multiple devices in parallel computing environments, which allows to operate on the data in parallel. For large models which cannot fit on single device, model parallelism turns out to be a good choice. Model parallelism (MP) \cite{bahdanau2014neural} partitions a DNN into disjoint subsets and trains each subset on a dedicated device, which reduces communication costs for synchronizing network parameters in a DNN but exposes limited parallelism as well as extra communication between model partitions.
\begin{figure}[t]
    \includegraphics[width=0.48\textwidth]{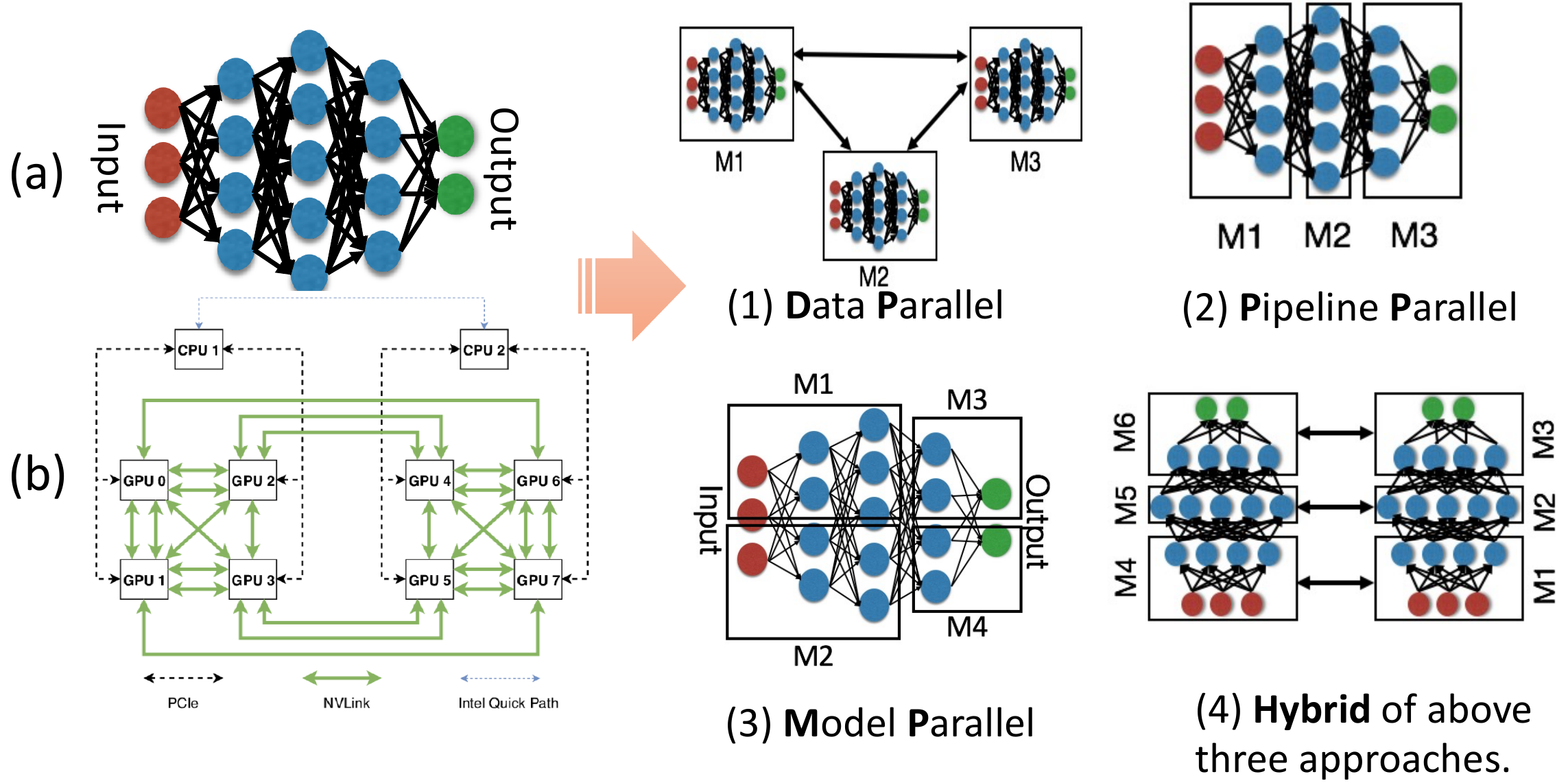}
    \caption{
     Typical types of parallelism.
    }
    \label{fig:dapple}
\end{figure}

\textbf{Pipeline parallelism} (PP) \cite{harlap2018pipedream, huang2019gpipe, fan2020dapple} goes beyond DP and MP, mixing inter-batch and intra-batch parallelism.
In pipeline scheme, one or more consecutive layers are grouped into stages and processed with separate GPU(s), and both the forward pass and backward pass of all the layers are scheduled in one stage. $Planner$\cite{narayanan2019pipedream, fan2020dapple} in PP is responsible for cutting model layers into stages and this approach improves device utilization through pipelining multiple micro-batches. Figure \ref{fig:dapple} shows the schematic of those three parallel strategies.

\textbf{Deep RL} has been proven to be successful with Deep Q-Learning (DQN)\cite{mnih2013playing} introducing the idea of using neural networks as a Q-function approximator. \textbf{Rainbow DQN} \cite{hessel2018rainbow} combining improvements in deep RL, and has been shown to be promising for further improvements of deep RL agents in benchmark environments. Although not so straightforward, We try to leverage rainbow agent to assist the $automatic$ search of massive distributed strategies space.

\begin{figure}[t]
    \includegraphics[width=0.5\textwidth]{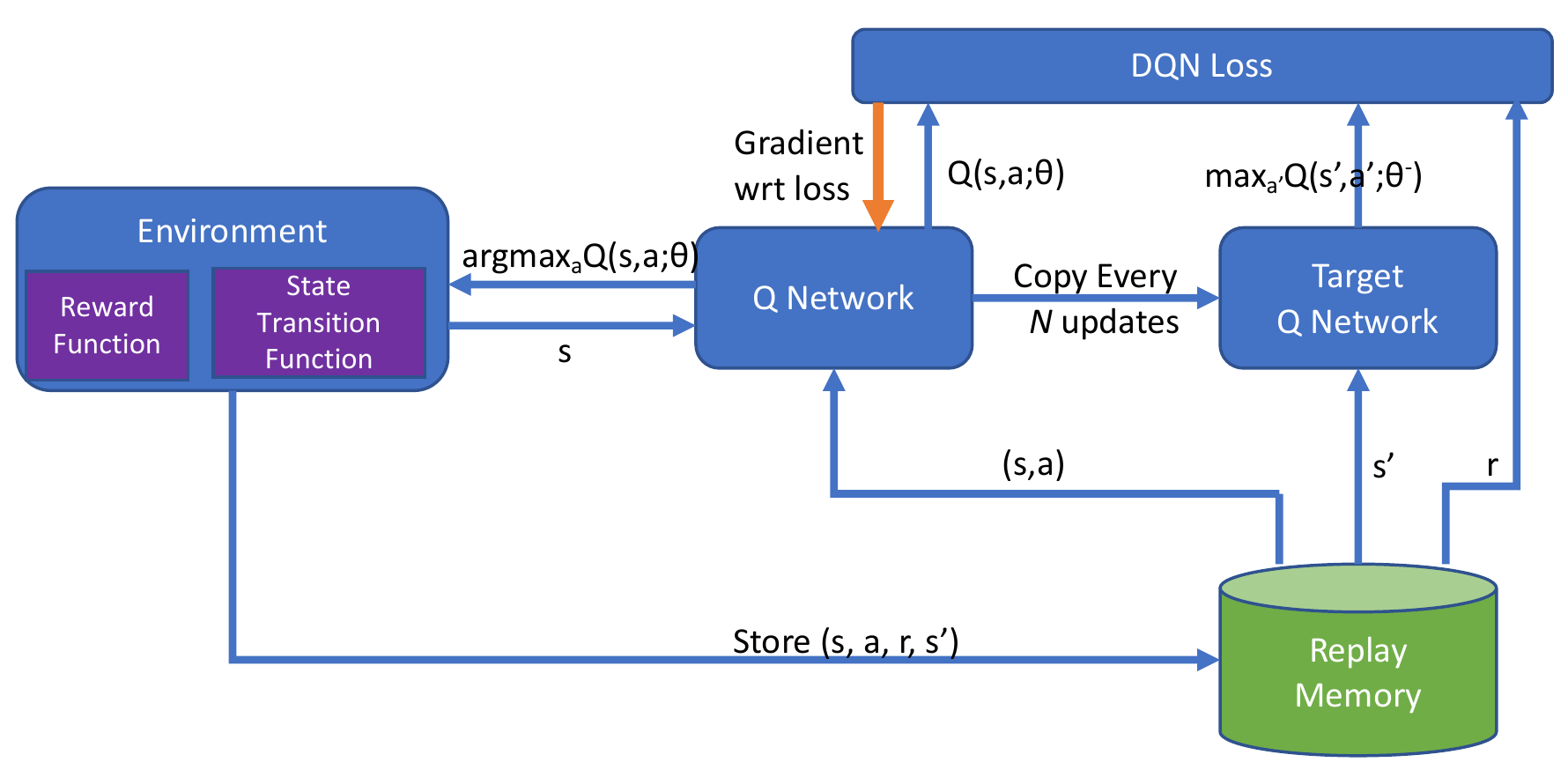}
    \caption{
     Workflow of leveraged $DQN$ method.
    }
    \label{fig:dqn-workflow}
\end{figure}

\textbf{The Rainbow agent.} Following the methodology from \cite{hessel2018rainbow}, we extend the
DQN algorithm with prioritized experience replay, double DQN, and dueling network architecture\cite{wang2016dueling}. Furthermore in contrast to \cite{hessel2018rainbow}, we apply the following changes to successfully train the Rainbow agent: (1) we discard the noisy linear layers \cite{fortunato2017noisy}, relying on $\epsilon$-greedy exploration instead. Since the agent was already required to learn environmental noises from the user simulator, a possible explanation could be that the inclusion of a second noise distribution might have been too difficult to learn. (2) We adjust the number of DNN layers for different tasks. As the greater the number of layers, the stronger the network learning ability.
Figure \ref{fig:dqn-workflow} shows the workflow of our leveraged $DQN$ method.
 
\textbf{Problem formulation for DQN algorithm on HLO IR.}
Formally, we define our learning task as follows. In reinforcement learning, the sequential decision-making problem is modeled using the Markov Decision Process formulation defined by the tuple $<S, A, R, S'>$.
For any Q-learning task, we need to define the following five aspects: state space, actions, rewards, policy and termination. 


We will illustrate our framework solving three optimization problems over directed HLO graphs. Let $G(V, E)$ denotes a directed HLO graph, where $V$ is the set of nodes, $E$ the set of edges. In our settings, each HLO instruction refers to one node of $V$ and the data-flow between producer instruction and consumer instruction refers to the corresponding edge of $V$. Specially, we refer the nodes with no inputs as $Source Nodes$, those nodes with no outputs as $Sink Nodes$ and the others as $Compute Nodes$.
Give device topology $D$, these optimization problems are:
\begin{itemize}
    \item \textbf{Auto Data Parallelism (ADP)}: Given a graph $G$, find a subset of dimensions of $Source nodes$ from $V$ such that communication overhead of the propagation graph from the selected slicing dimension of $Source Nodes$ to $Sink Nodes$ is minimized.
    \item \textbf{Operator Partitioning Parallelism (OPP)}: Given a graph $G$, find a slicing strategy of all dimensions of all $trainable\ variables$ of $G$, such that the average device utilization is maximized.
    \item \textbf{Pipeline Parallelism (PP)}: Given a graph $G$ and the number of stages ($K$) expected to be split, find a subset of nodes $S\subseteq V$ such that the pipeline length with cross-stage communication overlap considered is minimized.
\end{itemize}

\section{Auto-MAP Approach}

\begin{figure}[H]
    \includegraphics[width=0.45\textwidth]{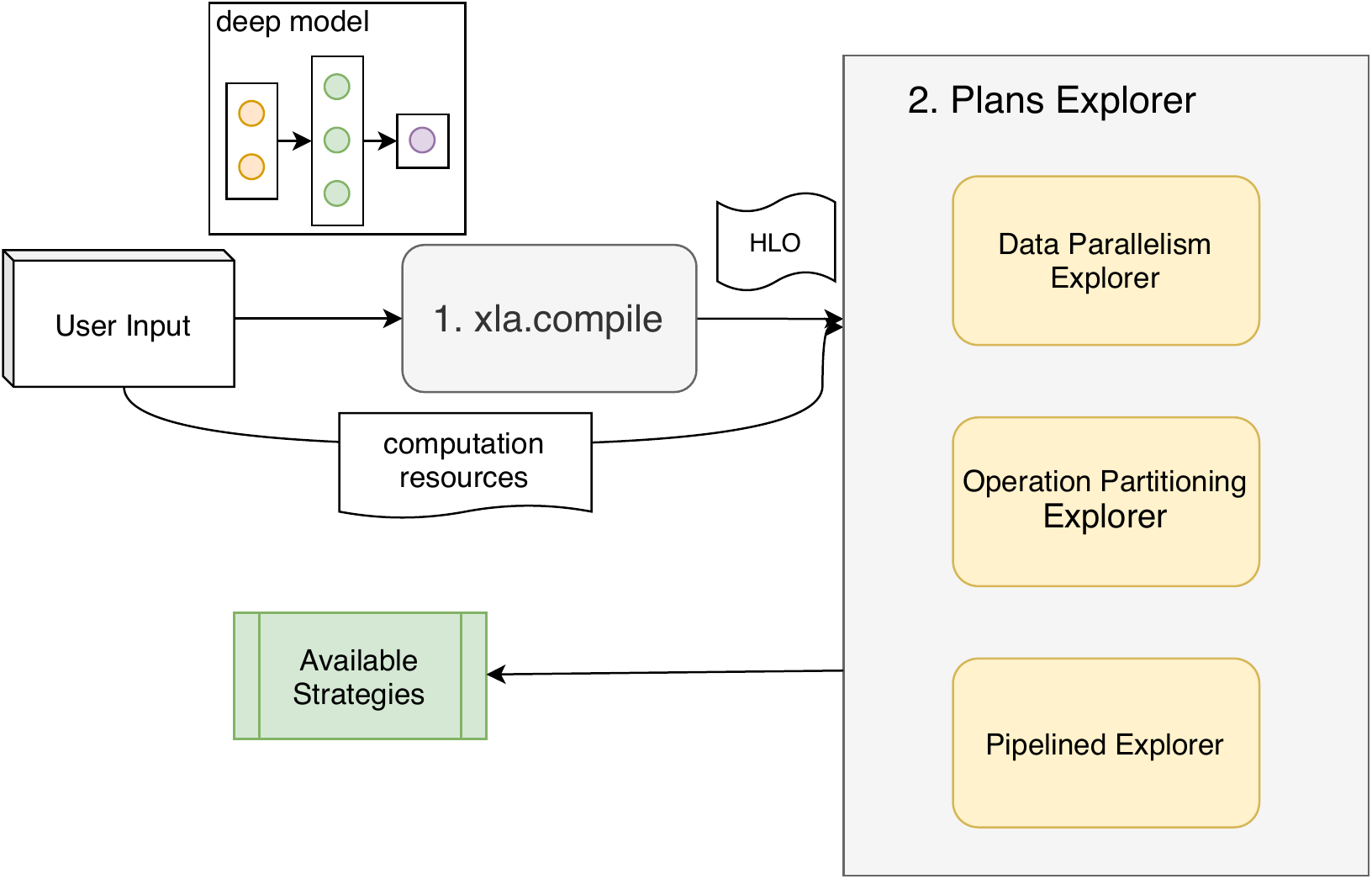}
    \caption{
     Workflow of our approach.
    }
    \label{fig:workflow}
\end{figure}

\subsection{Exploration Workflow}
In order to decouple the distributed plans from the APIs of specific deep learning framework, the exploration module should be constructed on an intermediate representation layer designed for describing computation flow of deep learning task. Specifically, we build our search algorithm over HLO borrowed from TensorFlow XLA. Figure \ref{fig:workflow} shows the workflow of our approach. Taken deep models written by any framework (e.g. TensorFlow, PyTorch, MXNet\cite{chen2015mxnet}), the XLA compiles and transfers the original flexible computation graph into HLO IR. The \emph{Plans Explorer} will search three different categories of plans including data parallelism, operator partitioning parallelism and pipeline parallelism over HLO based on given computation resources.

For pipeline parallelism, we only do cut on the forward computation subgraph in HLO, which can be detected by the meta information of each instruction. Both the online inference and online training approach are provided respectively to explore pipeline parallelism. Users need to specify the number of stages in advance for both approaches. Finally, the workflow produces the best one among all available candidate plans. 

We explore these three different categories of plans separately. To cope with the huge solution space and provide totally workload neutral plan, we use DQN approach combined with heuristic pruning instead of ordinary heuristic algorithms to search. For a specified workload, the corresponding solution would be found during training stage or inferred from models that has been trained offline. To adapt to the reinforcement learning training flow, \emph{state}, \emph{action} and \emph{reward} should be carefully designed according to their objectives. We briefly introduce our approach in the following subsections, and the details of design and implementations will be discussed in \ref{sec:impl}.

\subsection{Operator Partitioning Parallelism}
\subsubsection{DQN flow Setup}
Since the trend goes to increase the size of deep learning models, the on-device memory is a scarce resource for training tasks. Fortunately, the memory issue can be alleviated through model parallelism. In practice, an effective way to parallelize deep models is to partition operators, which not only alleviates the memory pressure but also parallelizes the computations. With operator partitioning, the saved memory can be used for injecting larger batch size to improve the GPU cluster utilization. 

Each instruction produces a unique variable in HLO. Therefore, to partition operators is identical to partition variables of each instruction. The derivation rules of each instruction are designed carefully for inferring partitioning decisions of unknown variables or parameters from the known ones. Obviously, some partitioning plans are invalid because some derivation rules are violated. This can only be detected during the procedure called \emph{propagation}, which performs derivation rules for each instruction when given the known partitioning decisions of variables or parameters. The propagation terminates when encountering the following three situations. (1) There is no enough information to derive the remains variables. (2) A conflict case is encountered for the violation of derivation rules. (3) All variables have been inferred without any conflict.  

We agree that only trainable variables which respect to model parameters may be partitioned in our approach. We also set the heuristic objective for operator partitioning parallelism to partition trainable variables as much as possible.

In Auto-MAP, since each trainable variable may has different dimension size, we make decisions for each dimension of each trainable variable about whether to be replicated or partitioned across all devices. These dimension status of all trainable variables are viewed as one strategy and the feasibility need to be verified by doing propagation in HLO.

\textbf{State and Action}. We define \emph{state} as one dimension vector which concatenates all dimensions partition status for trainable variables and there are three possible values at each position. And the \emph{action} is a binary flag which is True for partitioning across all devices and False for replicating among all devices. 

\textbf{Reward}. According to the objective mentioned above, we encourage partitioning by giving higher reward than replicating and punish the conflict case by giving negative reward. 

\subsubsection{Linkage Group}
\begin{figure}[t]
    \includegraphics[width=0.45\textwidth]{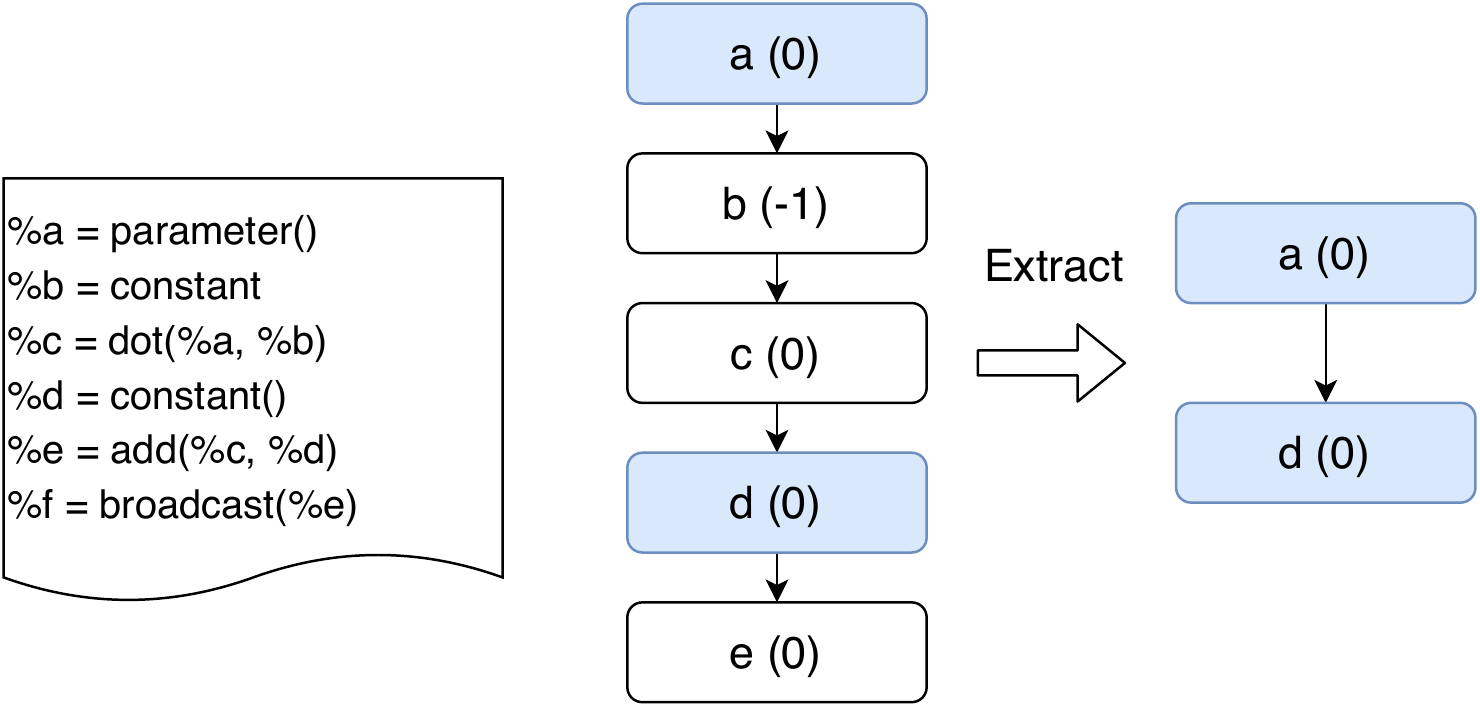}
    \caption{
      Linkage group example.
    }
    \label{fig:linkage_group}
\end{figure}

The search space of operator partitioning is so huge that even DQN requires lots of time based on the above setup, thus we introduce an heuristic pruning technique called linkage groups. Linkage group exists for each trainable variable, which records the deterministic partitioning decisions for other trainable variables caused by itself. Figure \ref{fig:linkage_group} illustrates the concept of linkage group. When the partition status of one dimension has been decided, the linkage group will be detected that whether current dimension with its partition exists. All the deterministic decisions of caused by current decision should be inferred via linkage group so that the search process can be greatly pruned to avoid unnecessary exploration. 

Due to the termination conditions of propagation mentioned above, the linkage group does contain only parts of partitioning decisions of other trainable variables. That is because the propagation procedure driven by one trainable variable with its decision always stops early when no enough information is given. However, larger linkage groups always perform better pruning effect.

\subsection{Auto Data Parallelism}
Implementing data parallelism over HLO is not intuitive because variables representing the input batch size cannot be easily identified. It is observed that the batch size dimensions will follow the data flow throughout the entire model. As a result, most variables expected to be influenced when the partition happens on the batch size dimension. With the help of propagation procedure, the variables represented training data and labels with their batch-size dimensions can be easily detected. 

More formally, the objective is to find the partition strategy for all input tensors, which results in the largest number of tensors to be partitioned. Moreover, the more tensors to be partitioned on the input tensor under the propagation rule, the closer to our objective. In Auto-MAP, the \emph{action} and \emph{reward} is almost the same compared to the operator partitioning task except the \emph{state}. Specifically, we define \emph{state} as one dimension vector which concatenates all dimensions partition status for all input tensors. 

\subsection{Pipeline Parallelism Exploration by Online Training}
\label{subsec:pipe_reward}

\begin{figure}[H]
    \includegraphics[width=0.45\textwidth]{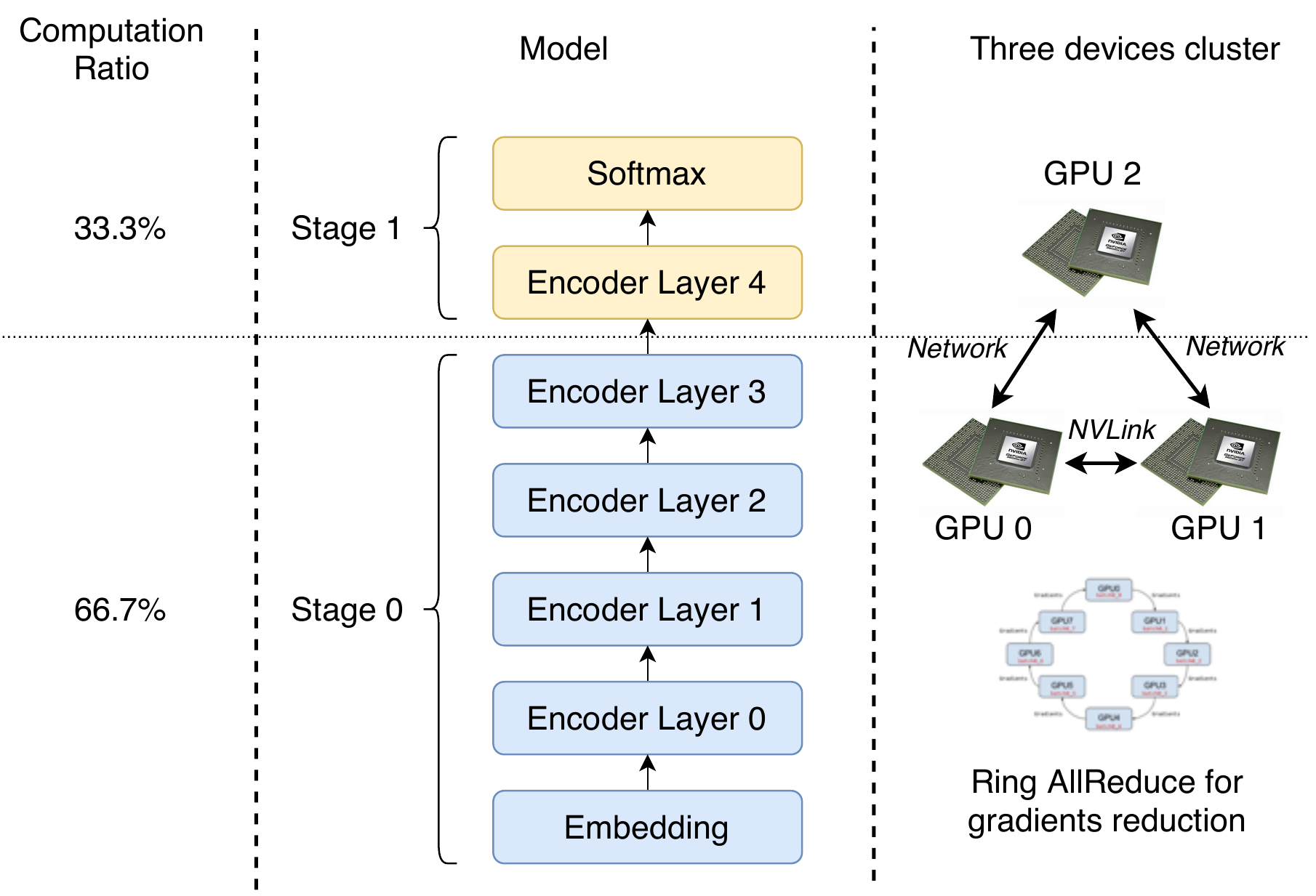}
    \caption{
      Cuts mapping from HLO to devices.
    }
    \label{fig:mapping}
\end{figure}

There are two key issues in pipeline partitioning. One is to cut the model into multiple stages, and the other is to place them onto a given GPU cluster. In industry, GPU clusters are always hierarchical which has relatively higher communication bandwidth within each node than across nodes\cite{DGX-1}. In Auto-MAP, we highlight that the exploration should be performed only on HLO. The main idea is that the distributed plan should allocate computation resources according to the computation ratio among all stages and the stage that allocated with more than one devices should be replicated. Figure \ref{fig:mapping} shows the common mapping from HLO to devices. Stage 0 is assigned with two devices with NVLink connection so that the gradients reduction could achieve good performance with NCCL\cite{nccl2019}. The activation between stage 0 and stage 1 are transmitted via Ethernet.

\textbf{State and Action}. Pipeline length is an effective way to estimate its performance and is influenced by the activation size across stage, the computation time and gradients all-reduce time in each stage. In Auto-MAP, we pre-compute these features at every possible pivot and encode them into one vector before applying the final cuts at the current step. The action outputs the pivot at each step. If one cut has been applied on HLO, the model will be further split into two stages and we limit the next cutting point should not happened at the previous stage.

\textbf{reward}. For a pipeline model, we can calculate pipeline length $L$ to estimate performance. In this case, we use $\frac{1}{\sqrt{L}}$ as reward for the higher performance could be achieved when $L$ is shorter.

\subsection{Pipeline Parallelism by Online Inference}

\subsubsection{Motivation} For Pipeline Parallelism Planning, we also present an alternative approach for a faster and generalizable way of inferencing an optimal hybrid parallelism strategy. This would allow us to train at a large generated dataset and inference on a real-world model. 

In order to find the optimal partitioning solution that yields maximal performance under our pipeline parallelism strategy, we need to 1) partition the model into different stages, 2) decide the replication factor for each stage, and lastly 3) map the stages to underlying hardware devices. In the following section, we will formulate our pipeline partitioning problem into a pure mathematical problem whose data can be randomly generated.

\subsubsection{Problem Formulation}
\label{sec:org68677af}

The original problem states: Given a/an HLO module \(H\), and the number of stages \(S\), find the optimal pipeline partition \(P\) that minimizes the end-to-end time of one-batch \(L\) with pipeline parallelism.

And with our profiler, we can get the per-instruction performance data \(C\), which represents the execution time on a given profiling device for each instruction in milliseconds. For communication, we use our DFG analyzer on \(H\) to calculate the parameter sizes \(W\) later used for allreduce time calculation, and activation communication \(A\) for each instruction if we partition the model at that specific instruction.

So this problem is now equivalent to: given three arrays of profiling data of an HLO model \(C\), \(A\) and \(W\), each of length the number of instruction in the original model \(H\), find a partition \(P\) that minimizes the end-to-end time \(L\), which we can calculate with our value function: \(L = V(P\ |\ C, A, W)\).

Since the number of instructions would certainly vary between models, and their profiling data might not even be close, we proceed with a round of data normalization described in the following section to ensure the training data has a consistent size and a reasonably close measure. And this this problem is now a array partitioning problem irrelevant to the input model, and the three arrays \(C\), \(A\), \(W\) can be generated on large scales.

Our first approach presented above uses DQN to search through the solution space of \(P\) for profiling data generated by each given model \(H\). This approach tries to train our DQN with generated data for this abstract array partitioning problem that could apply to real models upon inference.

\subsubsection{DQN Workflow}
\label{sec:orgbf6b496}

\textbf{State and Action}. 
We use the three performance metrics mentioned above ($C$, $A$ and $W$), and process the data along with device topology metrics to form the final state representation. The data processing will be detailed in section \ref{sec:impl}. 

\textbf{Reward}. 
Since we want to minimize the time of completing one global batch, we use $1/L$ as our reward.

\textbf{Training and inference}. First we use the data generation method detailed above to generate the training dataset, which will be then used to create a large number of environments ready to be interacted with. During the training process, since each environment represents a distribution of performance data, we will restrict the number of interactions with one environment to a very small number. In practice, we set each environment to be explored and exploited 50 times.

For testing, we used a freshly generated environment that is not in the training set, and evaluate its performance by letting the network inference the best partitioning solution, and assess its performance with our value function.

For real-world model inference, we do the same data pre-processing described in section \ref{sec:impl}, and output the best network inference result.


\section{Implementation}
\label{sec:impl}
\subsection{Overview}
All distributed execution plans can be unified into the same DQN workflow. In Auto-MAP, we select RAINBOW\cite{hessel2018rainbow} as the DQN framework built on PyTorch to go parallel search for all three category of strategies. We leverage cost model to estimate the performance of different plans so that the workflow can produce the best one among all candidates. 

The key issues of DQN workflow for different scenarios are environment, state, action and reward. We introduce our implementation of those for operator partitioning parallelism, auto data parallelism and pipeline parallelism, respectively.

\subsection{Operator Partitioning Parallelism}
\begin{figure}[t]
    \centering
    \includegraphics[width=0.30\textwidth]{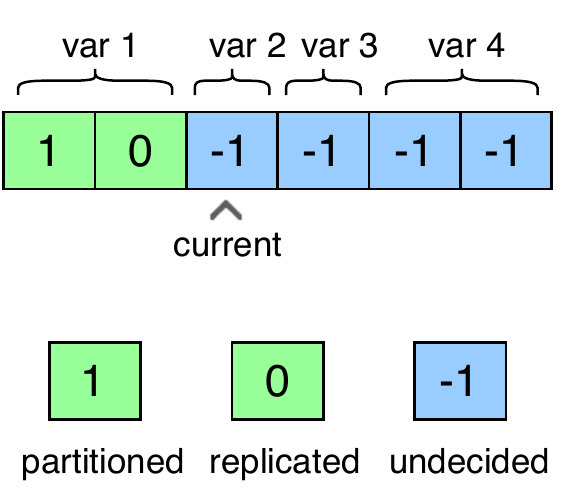}
    \caption{
      State representation in operator partitioning task.
    }
    \label{fig:operator_partitioning_state}
\end{figure}

\textbf{State and Action}. In our current implementation, the state contains a \emph{decision vector} and a\emph{current position}. Figure \ref{fig:operator_partitioning_state} shows the representation of decision vector. All dimensions of trainable variables are concatenated into an one dimensional vector. The 1, 0, -1 stands for partitioned, replicated and undecided status, respectively. The information of current position is an integer which indicates the index in the decision vector that will be decided in the next step. 

\begin{figure}[t]
    \centering
    \includegraphics[width=0.20\textwidth]{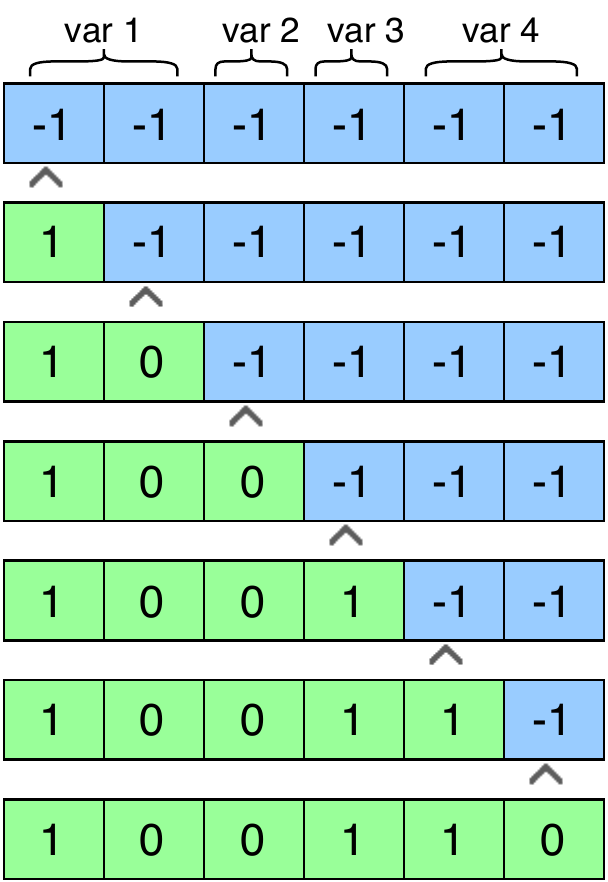}
    \caption{
      Partitioning variables in one episode.
    }
    \label{fig:progresses_one_episode}
\end{figure}

Initially, the decision vector is filled with all -1, which means no dimension is decided. Then, each dimension will be decided step by step in one episode until encountering an propagation conflict or all dimensions status have been decided safely. Figure \ref{fig:progresses_one_episode} shows one complete episode.

The action is implemented as a binary value, which the positive and negative represent to partition and to replicate, respectively, and the decision result will take effect on the current position. When all dimensions of one variable are marked with -1, it means this variable should be replicated across all devices.

\textbf{Reward}
We assign +0.4 and +0.1 reward to the case of partitioning and replication. A -1 reward will be given as the punishment when the conflict case is encountered caused by propagation in the entire HLO, while terminating current episode. 

\begin{figure}[H]
    \centering
    \includegraphics[width=0.48\textwidth]{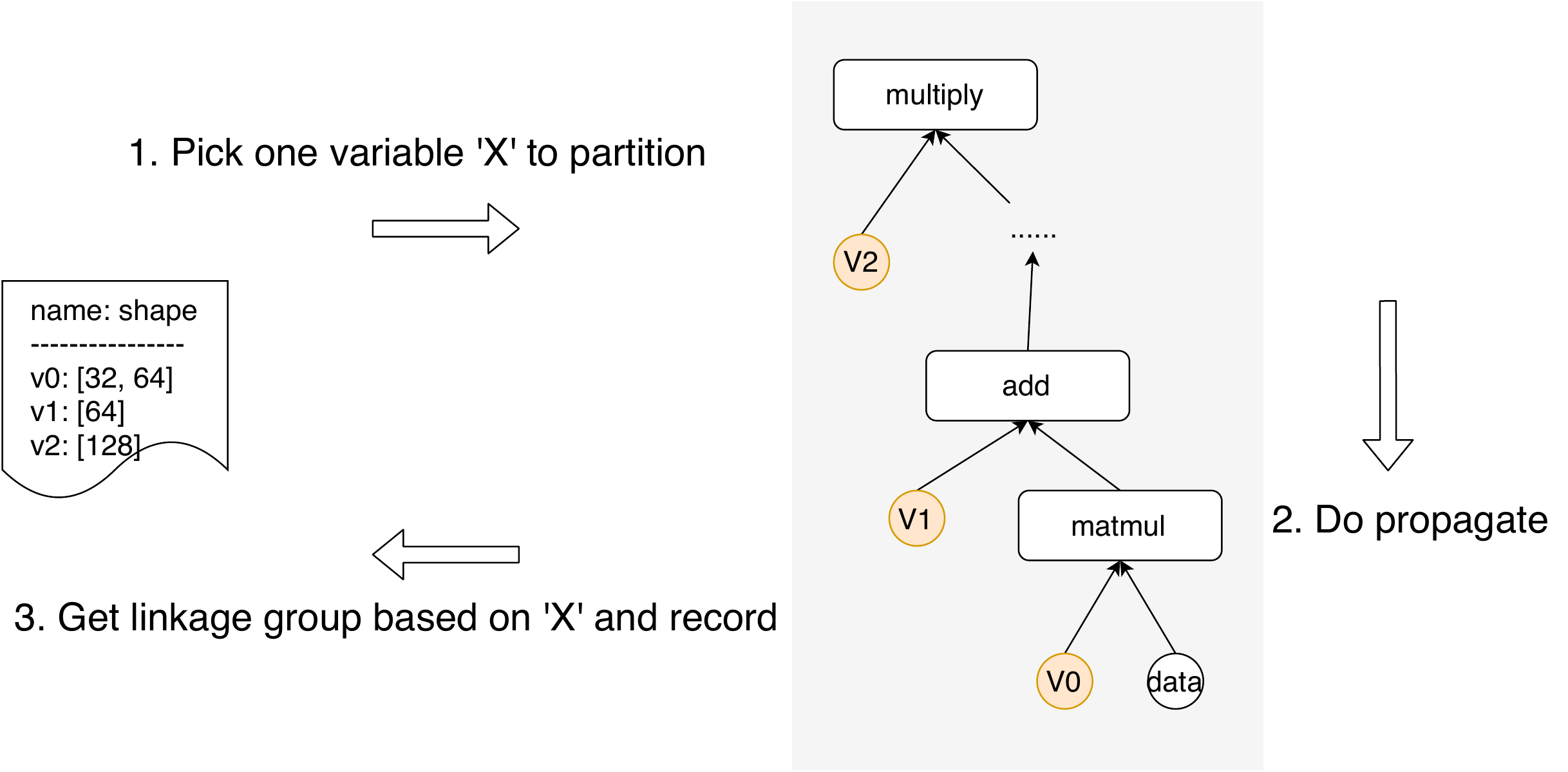}
    \caption{
      Linkage group extraction procedure.
    }
    \label{fig:linkage_extraction}
\end{figure}

\textbf{Linkage Group}
Linkage group should be extracted at the beginning of the DQN training task. The extracting procedure is displayed in figure \ref{fig:linkage_extraction}.

Linkage groups are formed by propagating each variable and its possible decision in the entire HLO. Specifically, we pick only one variable with its decision and send this pair into the propagation module to infer other variables' decision. Since propagation by only one variable and its decision cannot make deterministic decisions for every tensor, we only extract those deterministic ones. After all linkage groups have bee,  the decision order of every dimension in DQN task will be sorted according to the size of linkage group from large to small. 

With linkage groups, the reward is calculated according to the actual numbers of partitioned and replicated dimensions caused by current step if some decisions trigger more than one dimensions to be decided.

\subsection{Auto Data Parallelism}
\textbf{State and Action}. The philosophy of designing the state and action are the same with the case in operator partitioning parallelism. Since the trainable variables, hyper-parameters and training data with labels are all in the input list, we need to filter trainable variables and hyper-parameters out as much as possible. There is a heuristic that the constant tensors are definitely hyper-parameters and the trainable variables are marked outside HLO, so it is not difficult to find all possible candidate tensors.

We construct all candidate tensors into an one dimensional vector. Moreover, the current position index is also needed. And the action and reward are the same as we design in searching operator partitioning parallelism plan so that the Q network is guided to partition tensors as much as possible. This is always consistent with the reality that the greater the number we partition, the more intermediate tensors will be affected. Above all, the only difference is that we do not have any linkage groups. 

\textbf{Reward}. We use exactly the same reward as in operator partitioning parallelism problem for guiding the Q network to partition variables as much as possible.

\subsection{Pipeline Parallelism Exploration by Online Training}
\label{subsec:online_training}
In order to reasonably simplify the placement problem which maps from HLO-cuts to device-cuts, we treat the hierarchical computation topology as linear model which starts from the first device in first node. However, the search space is still huge and contains lots of solutions that are unnecessary to be explored. From a practical perspective, the solutions of better quality always happened when cutting on the pivots that exactly maps to the network boundaries or their nearby. Moreover, each stage contains at least one variable is also required in our implementation for the objective that to balance variables loading on different devices. We apply these two heuristic pruning methods to filter out some candidates pivots before training in our implementation. 

Firstly, we take the device-cuts which performs cutting on network boundary as \emph{center solution}. Then, A threshold number is specified as radius to represent the available range around each device-cut in the center solution. Thirdly, the device-cuts are filtered out according to the center solution and radius. Finally, All possible pivots which maps from the device-cuts will be left as our candidate pivots. 

\begin{figure}[t]
    \centering
    \includegraphics[width=0.45\textwidth]{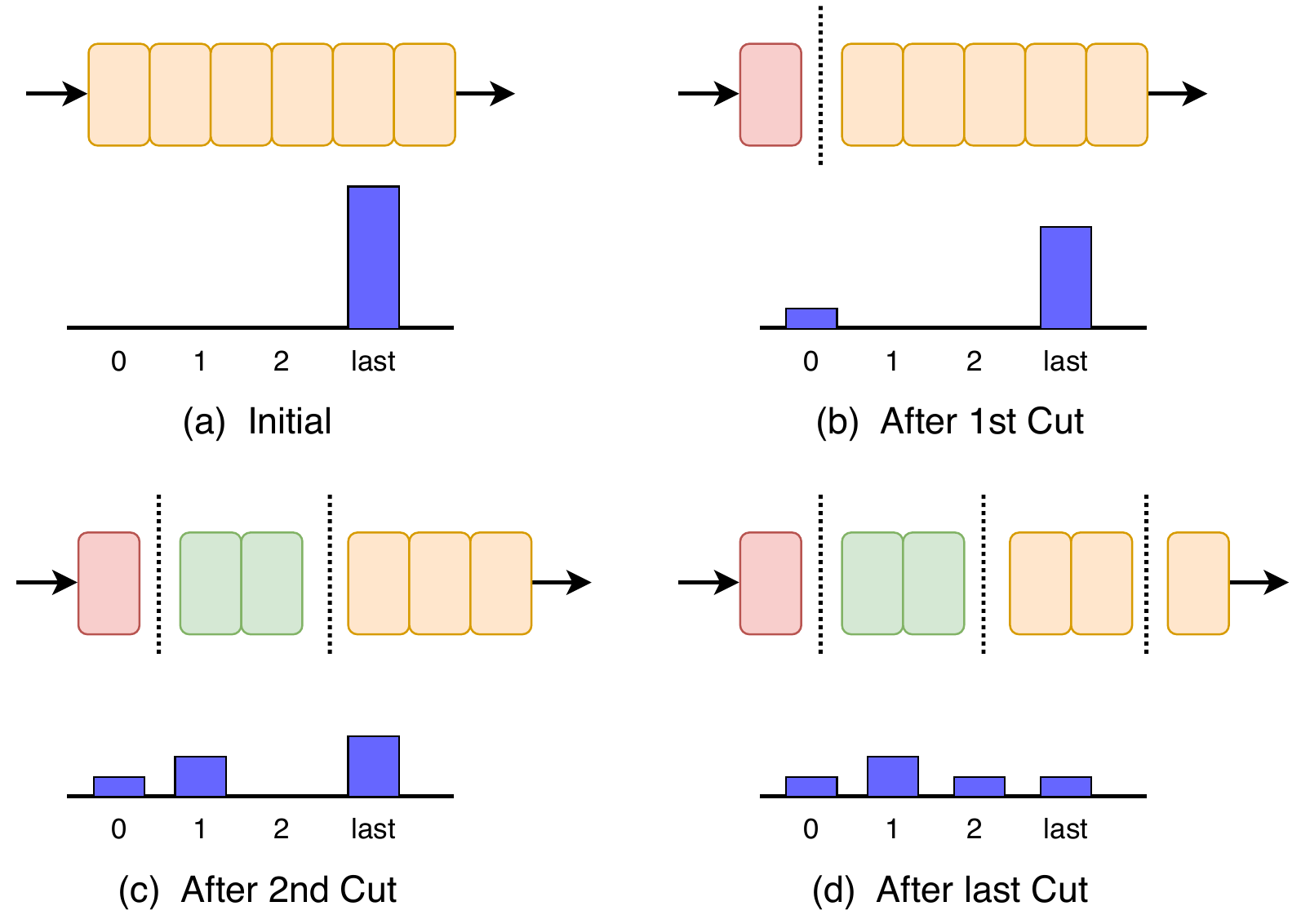}
    \caption{
      The change of AllReduce time cost for each stage when cutting deep model in one episode.
    }
    \label{fig:cutting_pro}
\end{figure}

\textbf{State and Action}. We pre-compute three features at each stage to encode state representation. We pre-compute the gradients all-reduce time of entire pipeline if we cut at any pivot in HLO at current step. This feature is very useful when there is a non-negligible bandwidth gap between devices within one node and cross nodes. The gradients reduction will be time-consuming when some stages cross nodes caused by cutting on the inappropriate position. Figure \ref{fig:cutting_pro} shows an example when cutting a deep model into four stages in one episode and the corresponding time cost of gradients reduction in each stage at each step.

The maximum activation transmission time is also required among all stages if we cut at any pivot at current step. To guide the time cost of each state towards balance, the computation balance ratios between minimum stage and maximum stage at any pivots are pre-computed.

Masking the unnecessary pivots is necessary when making cutting decisions on HLO. Actually, there are two kinds of pivots should be applied with mask. One is the pivot that we have filtered out in pruning stage, the other is the pivot that in an previous stage. In order to mask them on the output of Q network, we set their Q values to $-inf$ that represents the lowest expectation on that action. 

\textbf{Reward}. The pipeline length $L$ of a deep model can be calculated when given the pipeline parallelism plan by our cost model. As mentioned in \ref{subsec:pipe_reward}, the pipeline reward is designed as $\frac{1}{L}$. Moreover, the memory constraint should also be taken into consideration because some cutting strategy may encounter out of device memory. We give an $-\frac{1}{\sqrt{L}}$ to punish this case.

\subsection{Pipeline Parallelism Exploration by Online Inference}
We first describe data processing procedure, then introduce the DQN workflow.
\subsubsection{Data Processing}
\label{sec:org5f11c6c}

\begin{itemize}
\item Data Coarsening \& Normalization
\label{sec:org4f393ad}
Given a real-world model \(H\), we normalize the data into the same scale and size as data generated in the next section. This process is done in three steps: 1) building prefix sum, 2) coarsening array, and 3) normalization into [0, 1].

\begin{figure}[t]
    \centering
    \includegraphics[width=0.45\textwidth]{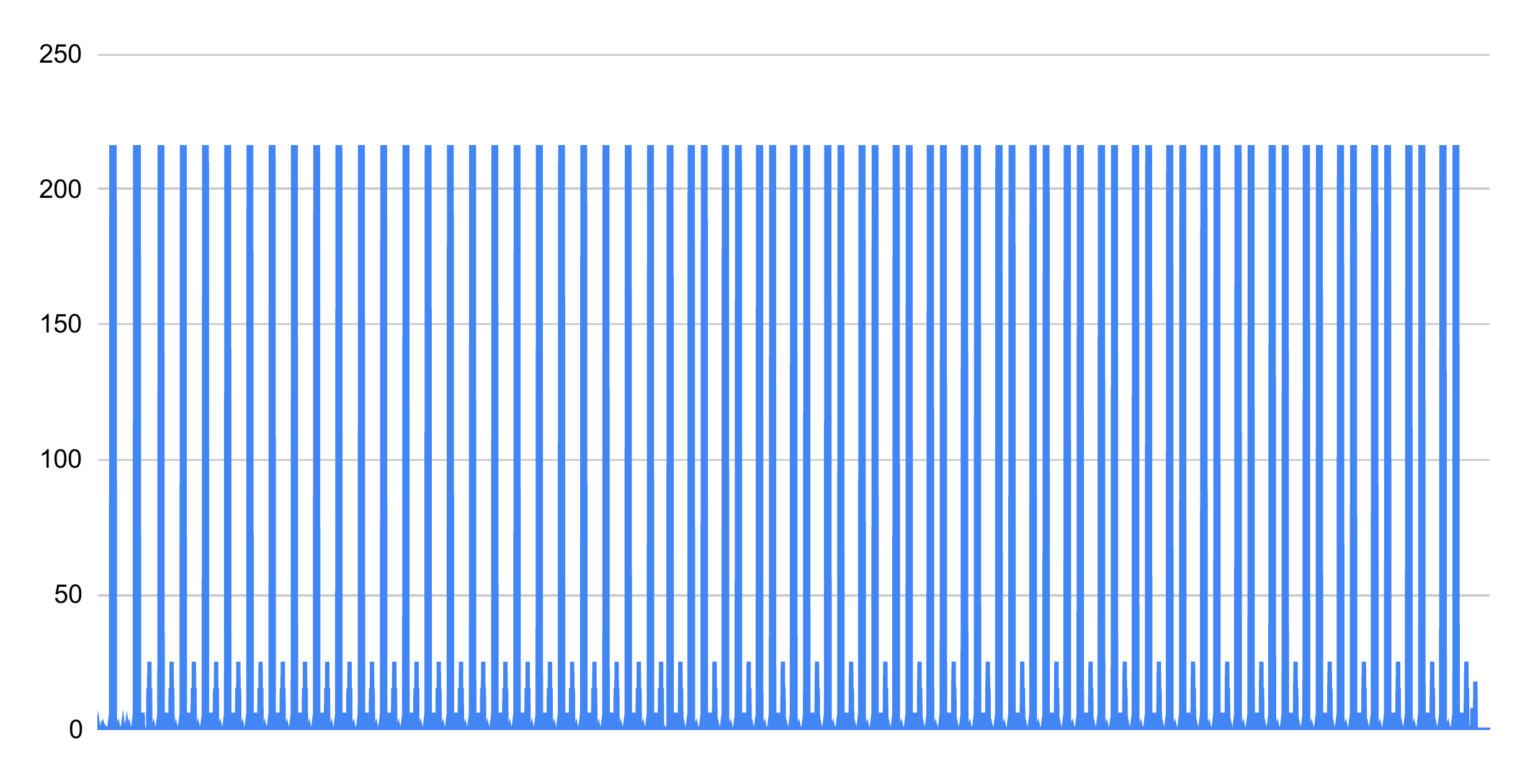}
    \caption{
      The input $C$ array, containing performance information of more than 50000 instructions.
    }
    \label{fig:data-1}
\end{figure}

\begin{itemize}
\item Step 1: Prefix Sum
\label{sec:org1b30198}
From profiling and DFG analysis on \(H\), we can get the profiling data \(C\), \(A\), and \(W\). We first build the prefix sum array for computational data \(C\) and parameter size \(W\): \(C' = prefix\_sum(C)\), \(W' = prefix\_sum(W)\).

The two updated array \(C'\) and \(W'\) accessed at index \(i\) now represents the computational time / AllReduce size of the first \(i\) instructions. \(A\) array is left untouched because it does not make sense to sum up all the cross-stage communication before a specific instruction. \(A[i]\) still represents the estimated cross-stage communication if the model was cut at instruction \(i\).

\begin{figure}[t]
    \centering
    \includegraphics[width=0.45\textwidth]{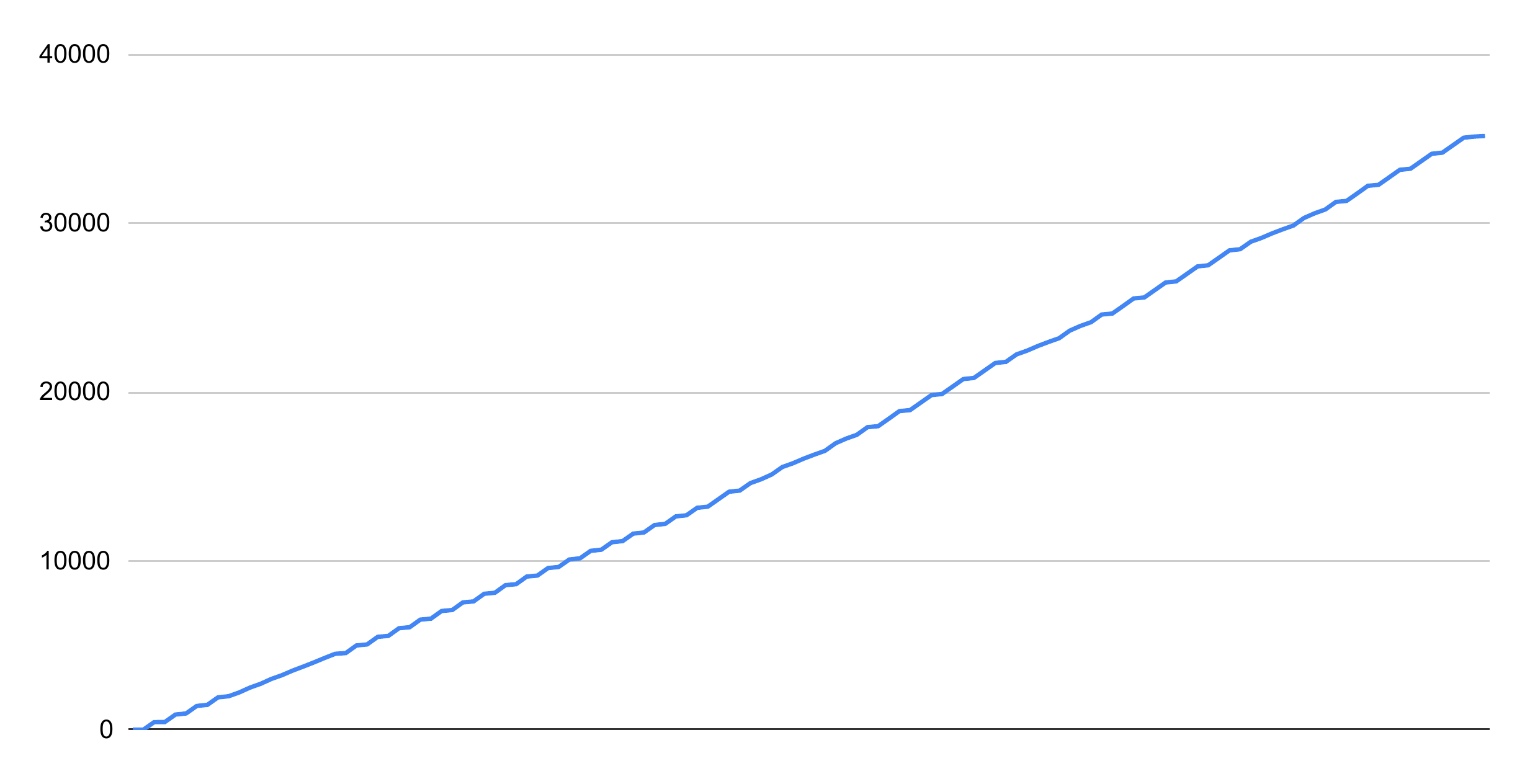}
    \caption{
      The input $C$ array, after building prefix sum.
    }
    \label{fig:data-3}
\end{figure}

\item Step 2: Coarsening
\label{sec:orgfc5257f}
In order to adapt to models of different sizes, we need to scale the profiling data to a fixed number, which we empirically set to 128. For the above three arrays, we evenly take 128 points to form the new arrays: \(C'', A'', W''\).

We can do this to \(C'\) and \(W'\) because they are already in prefix sum form, and \(A\) also because the cross-stage communication is specific to each instruction. After the coarsening, we lost the possibility to partition into instructions that are not in those 128 points, but the problem is now irrelevant to the input model size.

\begin{figure}[t]
    \centering
    \includegraphics[width=0.45\textwidth]{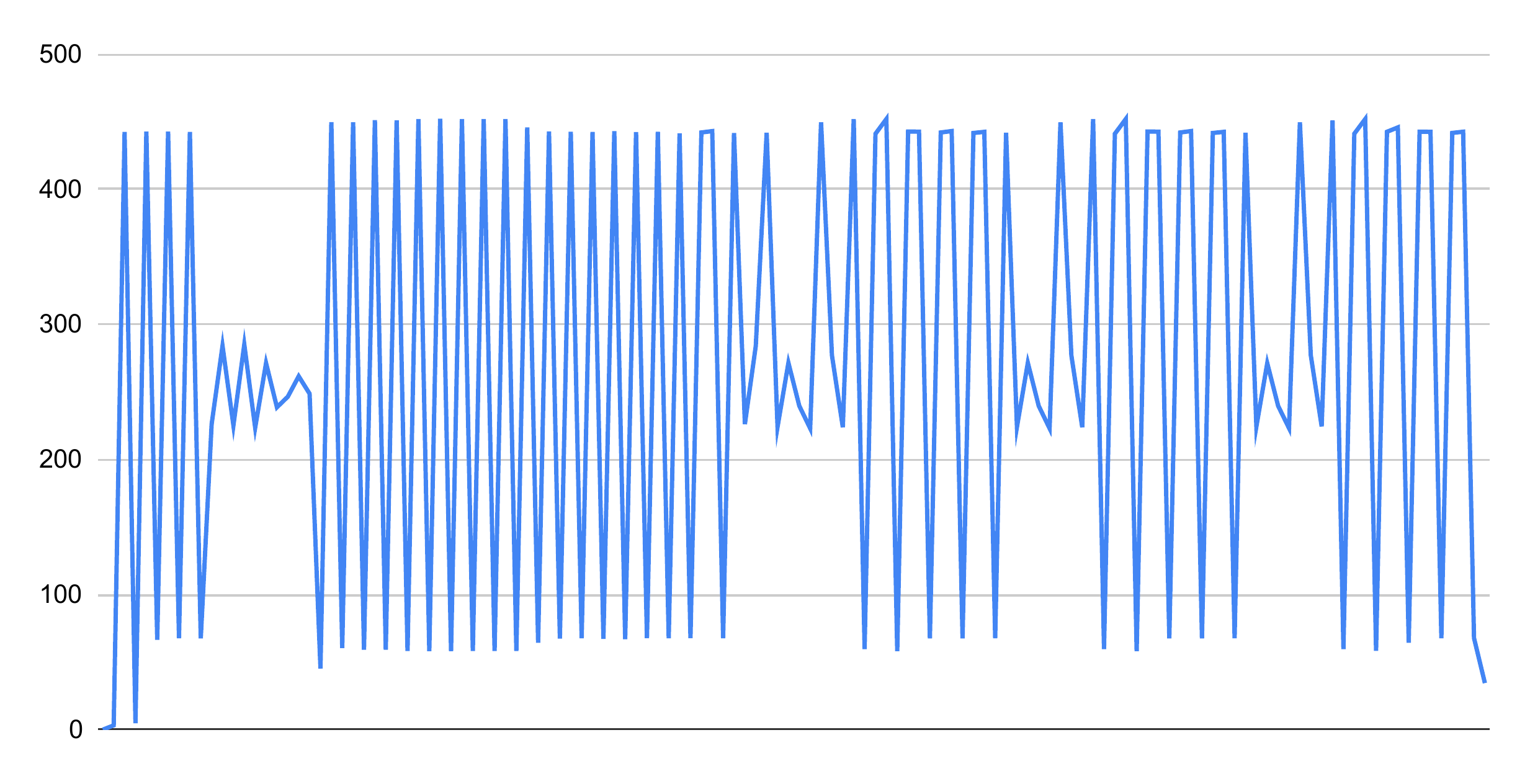}
    \caption{
      The $C$ prefix sum, coarsened to granularity 128.
    }
    \label{fig:data-2}
\end{figure}

\item Step 3: Normalization
\label{sec:org8b84af2}
Since we want to generalize across different models, and to generate large number of randomized data, some form of normalization is needed to keep all the data under a similar scale. In practice, we scale the three arrays simultaneously to \([0, 1]\): 
$$MAX = max(C'', A'', W'')$$
$$C^* = C'' / MAX$$
$$A^* = A'' / MAX$$
$$W^* = W'' / MAX$$

After this step, regardless of what the original model is, the resulting arrays \(C^*\), \(W^*\), \(A^*\) each has length 128, and the elements are all within \([0, 1]\). These three arrays essentially describe \textbf{the distribution of computational times, activation sizes, and parameters throughout the model in the time dimension}.
\end{itemize}

\begin{figure}[t]
    \centering
    \includegraphics[width=0.45\textwidth]{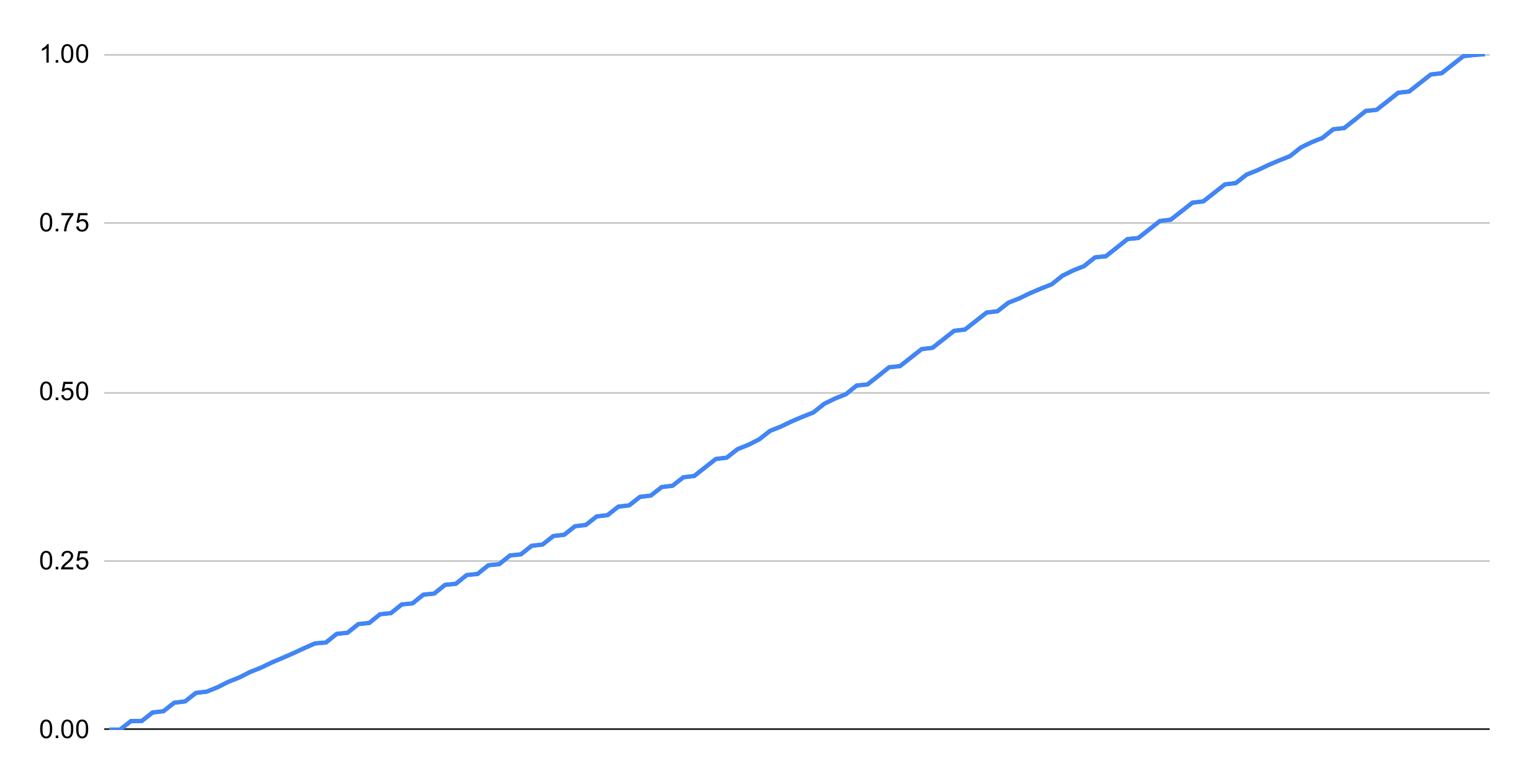}
    \caption{
      The $C$ 128-length prefix sum, renormalized to [0, 1].
    }
    \label{fig:data-4}
\end{figure}

\item Data Generation
\label{sec:orgf49c553}
To complement our existing model database, we choose to generate random data of different distributions that satisfies the requirement presented above. During the data generation, we use random number generator with optional distribution parameter (e.g. uniform, normal, binomial) to generate three arrays of float numbers ranging from 0 to 1, and then do the same transformation described above: build the prefix sum, coarsening and normalizing the array to get the generated \(C^*, A^*, W^*\).

In the actual training process, we will need to generate hundreds of thousands of these array groups to construct the training set and test set.
\end{itemize}

\subsubsection{DQN workflow}
\ \\

\textbf{State}. For our state representation, we have the following data fed into the network:

\begin{itemize}
\item Computational Times \(C^*\)
\item Activation Sizes \(A^*\)
\item AllReduce Sizes \(W^*\)
\item Device Topology (square matrix describing the interconnect speed between any to device)
\item Intermediate Partition
\end{itemize}

All of them are resized to one-dimension tensor, scaled to \([0, 1]\) and concatenated into one single array to form the state.

\textbf{Action}. For this approach, we consider both HLO partitioning and device assignment as actions, and they share the same action space. 

\textbf{Reward}. $$\frac{1}{L}$$ with \(L\) being the end-to-end training time for one batch, so that maximizing the reward means minimizing the end-to-end training time. This is the same as we used in \ref{subsec:online_training}.

\section{Experiments}
\label{sec:exp}
\subsection{Experimental Setup}

\paragraph{Benchmarks} We evaluate workloads for each distributed execution plan. Table \ref{tbl:benchmarks} summarizes all the five representative DNN models that we use as benchmarks in this section. 

\paragraph{HLO of workload}
We feed HLO Json files and trainable variable list of each workload as inputs into Auto-MAP framework. Another HLO text file is also provided
for debugging in our experiments.

\begin{table}[t]
    \caption{Benchmark models for each experiments.}
    \label{tbl:benchmarks}
    \begin{center}
    \begin{tabular}{lll}
    \toprule
    Task & Model & Params \\
    \midrule
    Language Model & BERT-48\cite{devlin2018bert} & 640M \\
    \cline{1-3}
    Machine Translation & \makecell[l]{T5-base\cite{raffel2019exploring} \\ T5-3B\cite{raffel2019exploring} \\ T5-11B\cite{raffel2019exploring}} & \makecell[l]{51M\\ 3B\\ 11B} \\
    \cline{1-3}
    Image Classification & VGG-19\cite{simonyan2014very} & 137M \\
    \bottomrule
    \end{tabular}
    \end{center}
\end{table}

\paragraph{Simulated Hardware Configurations}
Table \ref{tbl:cluster-spec} summarizes three hardware environments in our experiments. In our observation, the resources of 4 servers with 8 cards each are enough for training tasks. Therefore, we will give our execution plans with less than 4 servers.
\begin{table}[t]
    \caption{Simulated hardware configurations.}
    \label{tbl:cluster-spec}
    \begin{center}
    \begin{tabular}{ccccc}
    \toprule
    Config & Servers & \makecell[c]{GPU(s) per\\server($N_s$)} & \makecell[c]{Intra-server\\connnections} &\makecell[c]{Inter-server\\connections} \\
    \midrule
    A & 2 & 8x V100 & NVLink & 25 Gbps \\
    B & 3 & 8x V100 & NVLink & 25 Gbps \\
    C & 4 & 8x V100 & NVLink & 25 Gbps \\
    \bottomrule
    \end{tabular}
    \end{center}
\end{table}

\paragraph{Hyper-parameters} We fixed the training batch size to 64 and use the Adam optimizer\cite{kingma2014adam} with different initial learning rate to optimize different exploration tasks. For pipeline tasks, the initial learning rate would be set to 0.001. But for operator partitioning and auto data parallelism tasks, we set a smaller learning rate to 0.0005. 

As for the specific hyper-parameters in DQN, we fixed the $\gamma$ with 0.6 to all training tasks. We decay the exploration coefficient $\epsilon$ from 1.0 to 0.1 for all tasks, but the decay speed is totally different with respect to task type, which decay to the minimum after 2000, 500 and 10000 iterations for operator partitioning parallelism, auto data parallelism and pipeline parallelism, respectively.

Some general tricks for improving DQN convergence are also integrated in our training tasks. Specifically, we select the prioritized replay buffer\cite{schaul2015prioritized} and double DQN\cite{van2016deep} in rainbow and fixed the alpha and beta to 0.2 and 0.6 respectively. The frequency for updating target network is set to 100 and the replay buffer size is fixed to 2000 in all training tasks.

\subsection{Evaluation Results and Analysis}
\subsubsection{Operator Partitioning Parallelism}
There are already some partitioning strategies for transformer models\cite{shoeybi2019megatron}\cite{shazeer2018mesh}. It features to partition each attention block and the following MLP layer and all embedding variables while replicating other trainable variables, which is the same as the objective of Auto-MAP. For VGG-19, the effective way is to partition the last MLP block when given an hierarchical hardware configuration like Config B or Config C\cite{krizhevsky2014one}. Table \ref{tbl:t5_partition} and Table \ref{tbl:vgg19_partition} show our partitioning strategy of trainable variables for T5 family and VGG-19, where $-1$ means replication and the number greater than 0 represents the index of partitioned dimension. These partition strategies are consistent with our expectation. We have already known the ground truth of these workloads so that the quality of strategies could be measured in our experiments. We count the variables that should be partitioned as the target for each workload and to observe time cost to approach it. It is should be noted that some workloads need a finetuning stage to explore solutions of better quality.  

\begin{table}[t]
    \caption{T5 family partition results for each variable. We use -1 to represent to replicate the variable, and the positive number means the partition index of variable.}
    \label{tbl:t5_partition}
    \begin{center}
    \begin{tabular}{lc}
    \toprule
    Block or Layer & Variable Partition Strategy \\
    \midrule
    Self-attention & \{q=1, k=1, v=1, o=0\} \\
    \cline{1-2}
    MLP & \makecell[c]{\{conv1/kernel=1, conv1/bias=0, \\conv2/kernel=0, conv2/bias=-1\}} \\
    \cline{1-2}
    Embedding & \{ embedding\_weights=0 \} \\
    \cline{1-2}
    Layer normalization & \{scale=-1, bias=-1\} \\
    \bottomrule
    \end{tabular}
    \end{center}
\end{table}

\begin{table}[t]
    \caption{VGG-19 partition results for each variable . We use -1 to represent to replicate the variable, and the positive number means the partition index of variable.}
    \label{tbl:vgg19_partition}
    \begin{center}
    \begin{tabular}{lc}
    \toprule
    Block or Layer & Variable Partition Strategy \\
    \midrule
    Conv layers & -1 for all conv layers \\
    \cline{1-2}
    FC Layer $\time4$ & \makecell[c]{\{fc1/kernel=1, fc1/bias=0,\\fc2/kernel=0, fc2/bias=0\}}  \\
    \cline{1-2}
    Softmax Layer & {predictions/kernel=1, predictions/bias=0}\\
    \bottomrule
    \end{tabular}
    \end{center}
\end{table}

\begin{table}[t]
    \caption{The performance for searching OPP with config B and C. PC is short for partitioning count.}
    \label{tbl:opp_partition}
    \begin{center}
    \begin{tabular}{lccccc}
    \toprule
    Model & PC target & \makecell[c]{PC in \\ stage 1} & \makecell[c]{1st stage\\ time cost} & \makecell[c]{PC in \\ stage 2} & \makecell[c]{stage 2\\ time cost} \\
    \midrule
    VGG-19 & 38 & 5 & 30s  & -  & -\\
    T5-base & 111 & 111 & 0.5h & - & - \\
    T5-3B & 432 & 397 & 0.74h & 432 & 0.2h \\
    T5-11B & 432 & 386 & 1h & 432 & 0.45h \\
    \bottomrule
    \end{tabular}
    \end{center}
\end{table}

\begin{figure}[t]
    \centering
    \includegraphics[width=0.45\textwidth]{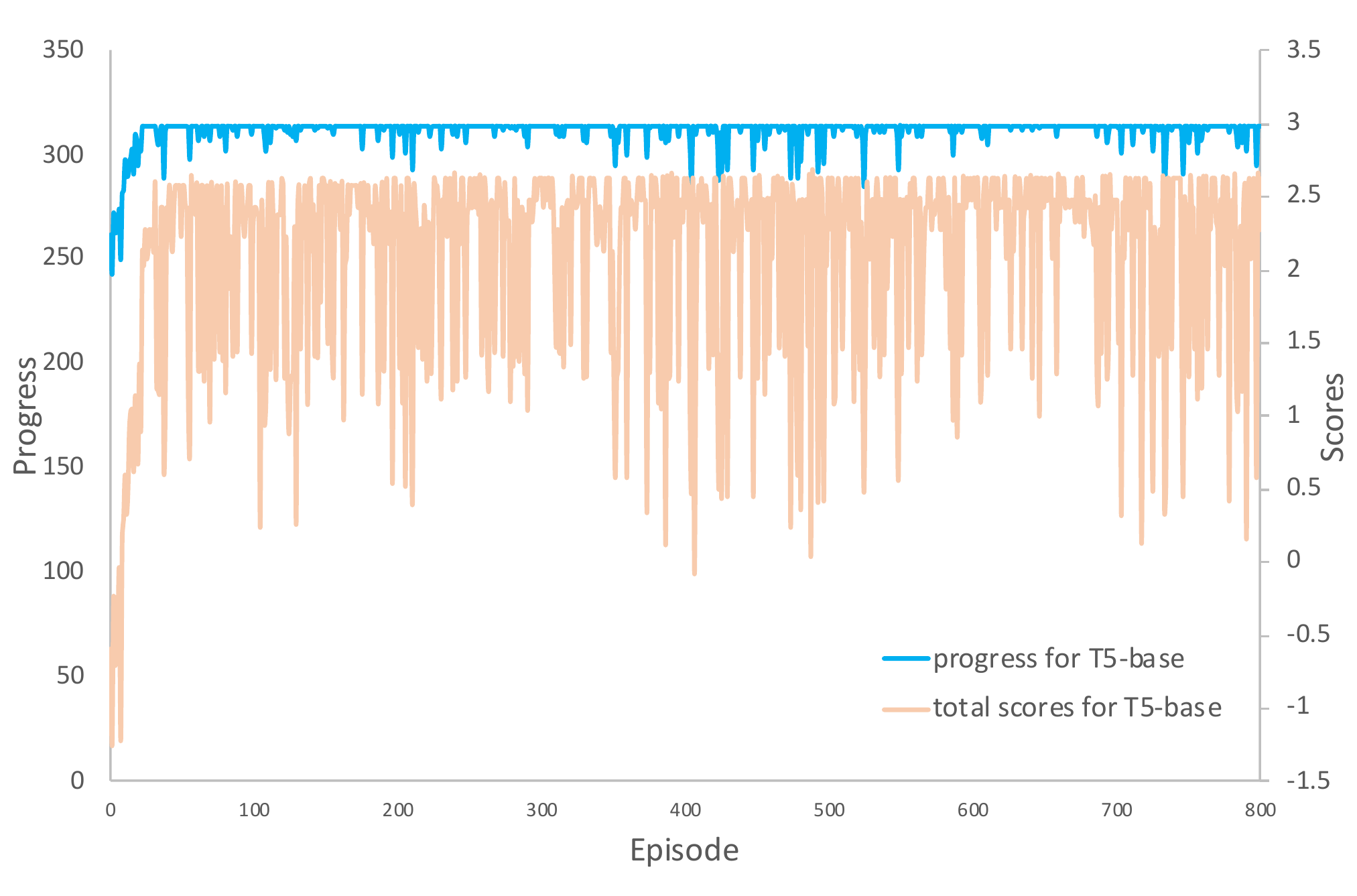}
    \caption{
      The convergence of T5-base in operator partitioning parallelism exploration task.
    }
    \label{fig:opp_convergence}
\end{figure}

We give the convergence for exploring operator partitioning parallelism on T5-base in figure \ref{fig:opp_convergence}. T5-base has 314 dimensions to be decided in total and 111 of them need to be partitioned according to the ground truth. With the help of linkage groups, DQN learns to avoid making conflicting decision quickly. It reaches the peak propagation progress and behaves more stable with higher scores as the time grows.

Table \ref{tbl:opp_partition} shows our searching performance on all benchmark models. We pay attention to the time cost to partition all variables which is required to be partitioned. We divide the search process into two stages. The first stage will search from scratch and may converge into a local minimum, while the second stage is to finetune from that result. Some workloads like VGG-19 and T5-base may not need finetuning stage mainly because the state space is relatively smaller than others, so it is easy to find the partition strategy as quickly as possible in first stage. However, some workloads like T5-3B and T5-11B with more trainable variables should involve with a finetuning stage. Specifically, when the partition strategy is stable in first stage, the program will stop current training phase and backtrace some variables which are marked with replication according to the linkage groups and start a finetuning stage. We found that even with the complicated case like T5-11B, the expected strategy could be found in two hours.

\textbf{VGG-19} As shown in Table \ref{tbl:vgg19_partition}, the solution that \emph{OPP} algorithm found is to replicate VGG-19's convolution layers while to partition the fully connected layers. This approach makes sense that for VGG-19 the last two FC layers occupy 86\% of the total parameters while the corresponding calculation time only accounts for 5\%. For such FC layer we prefer partitioning to replication for reducing gradients communication overhead in synchronous training. This desired distribution strategy as described above occurs in 30s (Table \ref{tbl:opp_partition}) while our DQN scores keep oscillating slowly and cannot converge quickly. One reasonable explanation is that our \emph{reward func} encourages splitting more variables while for 
VGG-19 is not the case as explained above. This implies that we need a more general reward function for models with different calculations and parameter distributions.

\textbf{T5-base}. The final solution is to split 111 variables and the partitioning results is the same with table \ref{tbl:t5_partition}. It is observed that our T5-base takes 0.5 hour to find the expected solution without a finetuning stage.

\textbf{T5-3B and T5-11B}. 3B and 11B has the same layers and variables counts but the variables size and the propagation time cost. The expected variables to be partitioned are 432 and the finetuning stages are required, which take 0.94 and 1.45 hour for 3B and 11B, respectively.
  
We infer that the DQN searching behaves better than enumeration. For example, T5-base has 188 trainable variables with at most two dimensions each, leads to a 376 binary vector which contains $2^{336}$ solutions in total and T5-3B and T5-11B contains $2^{1116}$ solutions to search. It is impossible for searching the expected solution within a limited time, while the DQN method could reach within 2 hours.

\subsubsection{Auto Data Parallelism}
We first filter out all trainable variables and constant tensors in input list in HLO IR to find the candidate tensors that possible to be training data.

T5-3B and T5-11B are not available for data parallelism for the memory issue. T5-3B needs at least two devices to load balance its variables and T5-11B consumes more devices, thus we display the results of T5-base in this part. Table \ref{tbl:batch_search} shows the results of auto data parallelism and all the tensor names in the table can be found in HLO text file. 

\textbf{VGG-19.} There are only 4 candidate tensors (with at most $4$ dimensions each) need to be partitioned for VGG-19 as shown in Table \ref{tbl:batch_search}. Our \emph{ADP} algorithm can converge steadily in $70$s to the first dimension of two tensors (namely arg0.1 and arg0.2). After manual verification, it is found that these two tensors are exactly the two inputs of the model: labels and features tensor respectively, and their first dimensions are exactly the \emph{batch size dimension} in the traditional sense.

\textbf{T5-base.} In our observation, we found that this procedure could be finished in half an hour. 
The search space is much less than in the operator partitioning problem. Specifically, there are 10 candidates with at most 4 dimensions each, leads to $2^{40}$ solutions. The DQN found the exact ground truth within 0.27 hour, while the enumeration would behave worse not only for the relative large solution space but also affected by the propagation time cost.

As the results shown in \ref{tbl:batch_search}, there are 7 tensors need to be partitioned in total and all of them choose to partition the first dimension, which is consistent with our intuition. In machine learning training task, we feed them with some sequence and other format data. The batch dimension is always at the first rank for each tensor.

\begin{table}[t]
    \caption{The experiment result of auto data parallelism. We use -1 to represent to replicate the variable, and the positive number means the partition index of variable.}
    \label{tbl:batch_search}
    \begin{center}
    \begin{tabular}{lclc}
    \toprule
    Model & Candidate count & Partition results & Time cost \\
    \midrule
    T5-base & 10 & \makecell[c]{\{arg0.1=0, arg1.2=0,\\ 
    arg2.3=0, arg3.4=0, \\ 
    arg12.13=0, \\arg17.18=0 \\
    arg22.23=0\}} &  0.27h \\
    \cline{1-4}
    VGG-19 & 4 & \makecell[c]{\{arg0.1=0, arg1.2=0\}}  & 70s  \\
    \bottomrule
    \end{tabular}
    \end{center}
\end{table}

\subsubsection{Pipeline Parallelism Exploration by Online Training}
We fixed all micro-batch sizes with 16 in all experiments. Then we do strategy search on Config A, B and C respectively. The number of stages to cut is depend on the number of servers under each hardware configuration. Both the strategy produced by online training and inference will be displayed.

In online training experiments, we set the center solution which performs the device-cuts on network boundary and set the radius to 3. Table \ref{tbl:pipeline_training} shows the online training experiments for searching pipeline parallelism. In order to make the experiments results human readable, we report not only the pivots cut on HLOs but also the corresponding layers nearby. Since each instruction in HLO produces a new tensor named with a $\%$ prefix, we display that tensor to indicates our HLO pivots. The device-cuts is displayed with an array which is filled with the cutting index of device. To cut the network boundary in the hierarchical topology hardware configuration like table \ref{tbl:cluster-spec}, the index should be the multiple of 8 because there are 8 cards within one server. 

We address that the time cost of DQN method is far better than enumeration, especially when we increase the stage number. That is mainly because each HLO contains at least thousands of instructions. Although we have filtered out some unexpected pivots and get a more concise candidates set, the search space is still large which costs more time by enumeration.

\begin{figure}[t]
    \centering
    \includegraphics[width=0.45\textwidth]{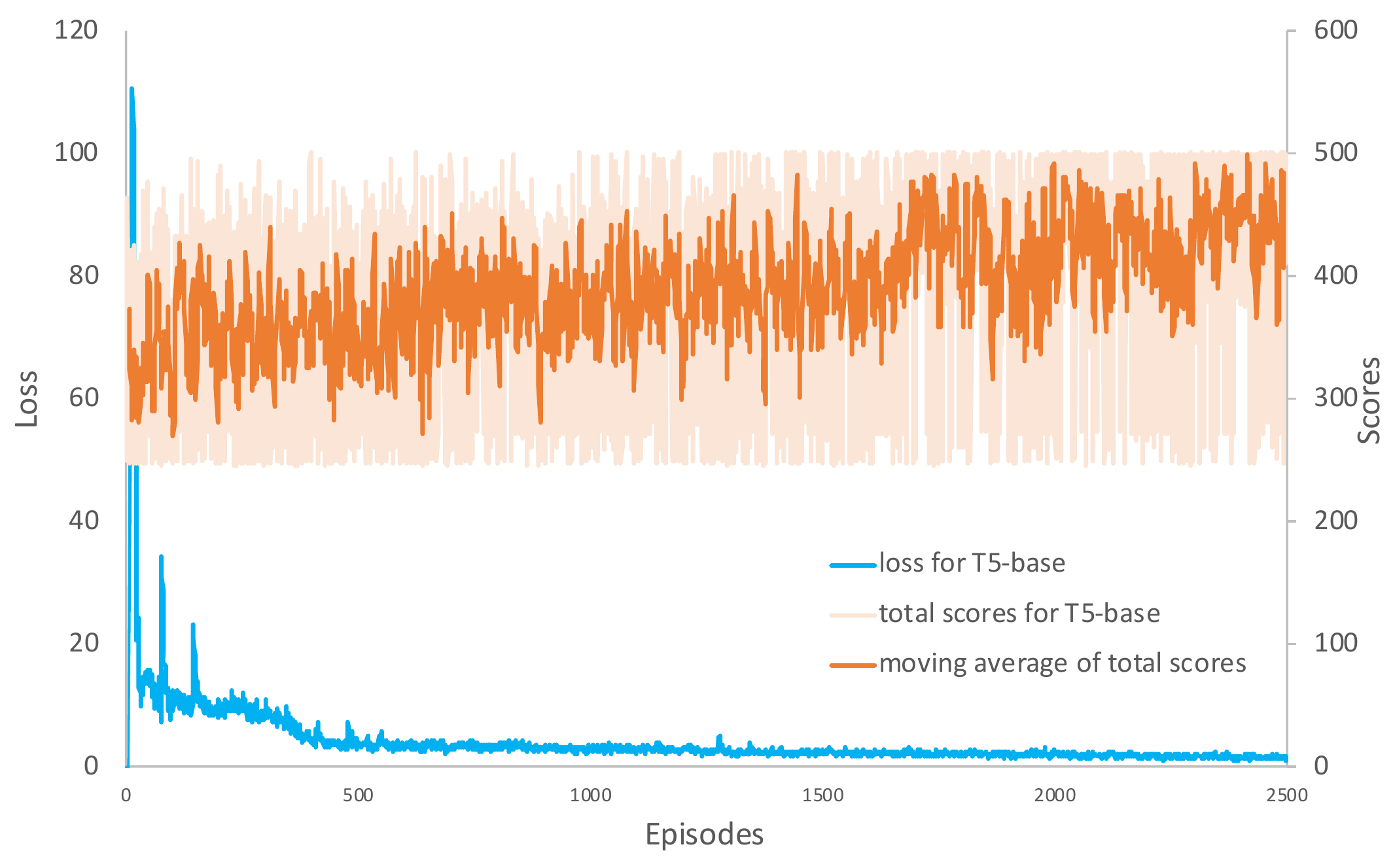}
    \caption{
      The convergence of T5-base in pipeline parallelism by online training task.
    }
    \label{fig:pp_convergence}
\end{figure}

We take the convergence of exploring pipeline parallelism by online training on T5-base as an example to show the training procedure with DQN. In figure \ref{fig:pp_convergence}, we The total scores is smoothed by applying moving average in order to show its trend. The figure shows that the loss drops very fast at the beginning and the trend of the total scores rises overall although the jitter is large. 

\begin{table*}[t] 
    \caption{The experiment result for searching pipeline parallelism by online training. }
    \label{tbl:pipeline_training}
    \begin{center}
    \begin{tabular}{llllll}
    \toprule
    Model & Config & Pivots on HLO & Corresponding Layer Nearby & Device cuts& Time cost\\
    \midrule
    Bert-48 & \makecell[c]{A \\ B \\ C} & \makecell[l]{(\%fusion.4004) \\ (\%dot.17952, \%fusion.3779) \\ (\%dot.16644, \%dot.22128, \%dot.27627)} & \makecell[l]{(layer23) \\ (layer17, layer32) \\ (layer14, layer27, layer40)} & \makecell[l]{(8) \\ (8, 16) \\ (8, 16, 24)} & \makecell[c]{95s \\ 157s \\ 262s} \\
    \cline{1-6}
    T5-base & \makecell[c]{A \\ B \\ C} & \makecell[l]{(\%fusion.1039) \\ (\%reshape.6314, \%transpose.12159) \\ (\%reshape.5622, \%fusion.1015, \%fusion.909)} & \makecell[l]{(dec/layer2) \\ (enc/layer5, dec/layer5) \\ (enc/layer4/conv1, \\dec/layer3, softmax)} & \makecell[l]{(8) \\ (8, 16) \\ (8, 16, 22)} & \makecell[c]{42s \\ 50s \\ 198s} \\
    \cline{1-6}
    T5-3B & \makecell[c]{A \\ B \\ C} & \makecell[l]{(\%dot.2204) \\ (\%reshape.13395, \%multiply.15030) \\ (\%reshape.13379,\\ \%multiply.15009, \%multiply.15049)} & \makecell[l]{(dec/layer2) \\ (enc/layer19, dec/layer11) \\ (enc/layer14, \\dec/layer5, dec/layer17)} & \makecell[l]{(8) \\ (8, 16) \\ (8, 16, 24)} & \makecell[c]{115s \\ 202s \\ 262s} \\
    \cline{1-6}
    T5-11B & \makecell[c]{A \\ B \\ C} & \makecell[l]{- \\ - \\ (\%fusion.4585, \%fusion.4241, \%dot.40105)} & \makecell[l]{- \\ - \\ (enc/layer13, dec/layer3, dec/layer15)} & \makecell[l]{- \\ - \\ (8, 16, 24)} & \makecell[c]{- \\ - \\ 280s} \\
    \bottomrule
    \end{tabular}
    \end{center}
\end{table*}

\textbf{Bert-48 and T5-3B}. The two models are very similar from the results. All strategies on \emph{Config A}, \emph{B} and \emph{C} proved that the cutting should be happened on the network boundaries, which are consistent with our expectation. Moreover, the pivots mapping to corresponding layer lead to almost uniform stages so that the computation on each stage are balanced. It takes about no more than 5 minutes to find them all.

\textbf{T5-base}. The strategies on \emph{Config A} and \emph{B} are similar with the case of Bert-48 and T5-3B. The time cost to converge is no more than 4 minutes. The strategy on \emph{Config C} is different for the last cut happens on the 22th device, which is a NVLink boundary. This is because the constraint that each stage contains one trainable variable at least. T5-model is too small for cutting 4 stages that the last cut should not happen beyond the 22th index. 

\textbf{T5-11B}. This model is huge enough so that it will cause OOM if it is cut less than 4 stages. Therefore, the DQN cannot find even one available strategy on \emph{Config A} and \emph{Config B}. For \emph{Config C}, the result is consistent with our expectation for cutting on the network boundaries of device topology. The time cost order of magnitude is the same with other models.

\subsubsection{Pipeline Parallelism Exploration by Online Inference}
Here we also present the results of our online inference approach. We trained our for $10^7$ episodes of environments constructed by random number generated with uniform and normal distribution. Our model is able to output the best hybrid parallelism solution for the NLP family models like BERT and Transformer-11B. 
For the CNN family, we need to finetune the model with the corresponding distribution for those models for another $10^4$ episodes before the model could correctly inference the best pipeline partitioning.

The detailed parallelism plan is presented in Table \ref{tbl:pipeline_inference}. 

\begin{table*}[t] 
    \caption{The experiment result for searching pipeline parallelism by online inference. }
    \label{tbl:pipeline_inference}
    \begin{center}
    \begin{tabular}{llllll}
    \toprule
    Model & Config & Partition Boundary & Corresponding Layer Nearby & Device cuts \\
    \midrule
    BERT-48 & \makecell[c]{C} & \makecell[l]{(34, 66, 98), granularity=128} & \makecell[l]{(layer14, layer27, layer40)} & \makecell[l]{(8, 16, 24)} \\
    \cline{1-5}
    T5-base & \makecell[c]{C} & \makecell[l]{(34, 66, 98), granularity=128} & \makecell[l]{(enc/layer4/conv1, dec/layer3, softmax)} & \makecell[l]{(8, 16, 24)} \\
    \cline{1-5}
    T5-11B & \makecell[c]{C} & \makecell[l]{(34, 66, 98), granularity=128} & 
    \makecell[l]{(enc/layer13, dec/layer3, dec/layer15)} & \makecell[l]{(8, 16, 24)} \\
    \bottomrule
    \end{tabular}
    \end{center}
\end{table*}

\section{related works}
Large DNN models are increasingly computational intensive and seriously consumption on device memory. It is a common practice to parallelize training
by leveraging multiple GPUs\cite{pal2019optimizing,jia2018exploring}.
Data parallelism, operator partitioning parallelism and pipeline parallelism are common approaches for distributed training
of DNN models.

\textbf{Auto Data Parallelism}. There are some high level frameworks aim at reducing the burden of users to automatically parallelizeing deep models using data parallelism\cite{Cheng2017TensorFlow}. 

\textbf{Operator Partitioning Parallelism}. For NLP models with attention blocks, some heuristic operator partitioning approaches\cite{shoeybi2019megatron, shazeer2018mesh} have already been proposed in recent years. For some convolutional networks like VGG-19 and AlexNet, it is a common practice to partition the last linear layers\cite{krizhevsky2014one, jia2018exploring}.

Some prior works and studies\cite{jia2018exploring, Jia2018Beyond}  focus on finding optimal distribution strategies over DNN layers.

\textbf{Pipeline Parallelism}. \cite{harlap2018pipedream,zhan2019pipe,huang2019gpipe,geng2019elasticpipe,yang2019pipemare}
has been proposed to train DNN by pipelining DNN models.
GPipe\cite{huang2019gpipe} explores synchronous pipeline approach to train large models while PipeDream\cite{harlap2018pipedream} explores the hybrid approach of data and pipeline parallelism
for asynchronous training.
The RL approach has been proposed to find optimal placement strategy for a given DNN\cite{Goldie2020Placement}.

\textbf{Rainbow DQN.}
Reinforcement learning (RL) is a general framework where agents learn to perform actions in an environment so as to maximize a reward.
DQN\cite{mnih2015human} is a RL algorithm that combines Q-Learning with deep neural networks to let RL work for complex, high-dimensional environments, like video games, or robotics.
Double DQN\cite{van2016deep}, Dueling DQN\cite{wang2016dueling}, Noisy DQN\cite{fortunato2017noisy} and DQN with Prioritized Experience Replay\cite{schaul2015prioritized} are these four important supplements which each of them handle a different aspect of an agent.
Rainbow DQN\cite{hessel2018rainbow} is an off-policy deep reinforcement learning algorithm that is the state-of-the-art technique in the field of reinforcement learning.
\section{conclusion}
\subsection{Summary}
We introduce Auto-MAP, a framework for exploring distribution strategies based on model architectures, which works on HLO IR and automatically discovers fast parallelization strategies with optimized DQN algorithm.
Data parallelism, operation partitioning parallelism and pipelined parallelism are all included in the exploration space. We leverage  DQN  with  task-specific  pruning  strategies to  help  efficiently  explore the search  space  including optimized strategies. Auto-MAP fully simplifies  the  user’s  burden  in  the  selection and  implementation  of  distribution  strategies. Our experiments show that Auto-MAP can find the optimal solution within two hours while achieving better throughput on several NLP and convolution models.

\subsection{Future Work}
Combination of HLO IR and DQN algorithm show convincing convergence results and performance. There are still some interesting works to follow.
First of all, replacing discrete DQN states with continues one for operation partitioning task for better interpretation and convergence. Secondly, currently our Auto-MAP framework can only give a single parallelization strategy automatically (i.e., DP, PP, operation partitioning), which may result in sub-optimal runtime performance in large-scale distributed training. In the future we will support exploring hybrid of these three strategies automatically. \emph{Auto-MAP} is open-source and will be made available to the public.

\bibliographystyle{IEEEtran}
\bibliography{./ref}

\begin{thebibliography}{10}
\providecommand{\url}[1]{#1}
\csname url@samestyle\endcsname
\providecommand{\newblock}{\relax}
\providecommand{\bibinfo}[2]{#2}
\providecommand{\BIBentrySTDinterwordspacing}{\spaceskip=0pt\relax}
\providecommand{\BIBentryALTinterwordstretchfactor}{4}
\providecommand{\BIBentryALTinterwordspacing}{\spaceskip=\fontdimen2\font plus
\BIBentryALTinterwordstretchfactor\fontdimen3\font minus
  \fontdimen4\font\relax}
\providecommand{\BIBforeignlanguage}[2]{{%
\expandafter\ifx\csname l@#1\endcsname\relax
\typeout{** WARNING: IEEEtran.bst: No hyphenation pattern has been}%
\typeout{** loaded for the language `#1'. Using the pattern for}%
\typeout{** the default language instead.}%
\else
\language=\csname l@#1\endcsname
\fi
#2}}
\providecommand{\BIBdecl}{\relax}
\BIBdecl

\bibitem{AI-and-compute}
D.~H.~D. Amodei, \emph{AI-and-compute}, 2019,
  \url{https://openai.com/blog/ai-and-compute/}.

\bibitem{rajbhandari2019zero}
S.~Rajbhandari, J.~Rasley, O.~Ruwase, and Y.~He, ``Zero: Memory optimization
  towards training a trillion parameter models,'' \emph{arXiv preprint
  arXiv:1910.02054}, 2019.

\bibitem{shoeybi2019megatron}
M.~Shoeybi, M.~Patwary, R.~Puri, P.~LeGresley, J.~Casper, and B.~Catanzaro,
  ``Megatron-lm: Training multi-billion parameter language models using gpu
  model parallelism,'' \emph{arXiv preprint arXiv:1909.08053}, 2019.

\bibitem{shazeer2018mesh}
N.~Shazeer, Y.~Cheng, N.~Parmar, D.~Tran, A.~Vaswani, P.~Koanantakool,
  P.~Hawkins, H.~Lee, M.~Hong, C.~Young \emph{et~al.}, ``Mesh-tensorflow: Deep
  learning for supercomputers,'' in \emph{Advances in Neural Information
  Processing Systems}, 2018, pp. 10\,414--10\,423.

\bibitem{jia2018exploring}
Z.~Jia, S.~Lin, C.~R. Qi, and A.~Aiken, ``Exploring the hidden dimension in
  accelerating convolutional neural networks,'' 2018.

\bibitem{geng2019horizontal}
J.~Geng, D.~Li, and S.~Wang, ``Horizontal or vertical? a hybrid approach to
  large-scale distributed machine learning,'' in \emph{Proceedings of the 10th
  Workshop on Scientific Cloud Computing}, 2019, pp. 1--4.

\bibitem{dryden2019channel}
N.~Dryden, N.~Maruyama, T.~Moon, T.~Benson, M.~Snir, and B.~Van~Essen,
  ``Channel and filter parallelism for large-scale cnn training,'' in
  \emph{Proceedings of the International Conference for High Performance
  Computing, Networking, Storage and Analysis}, 2019, pp. 1--20.

\bibitem{lepikhin2020gshard}
D.~Lepikhin, H.~Lee, Y.~Xu, D.~Chen, O.~Firat, Y.~Huang, M.~Krikun, N.~Shazeer,
  and Z.~Chen, ``Gshard: Scaling giant models with conditional computation and
  automatic sharding,'' \emph{arXiv preprint arXiv:2006.16668}, 2020.

\bibitem{huang2019gpipe}
Y.~Huang, Y.~Cheng, A.~Bapna, O.~Firat, D.~Chen, M.~Chen, H.~Lee, J.~Ngiam,
  Q.~V. Le, Y.~Wu \emph{et~al.}, ``Gpipe: Efficient training of giant neural
  networks using pipeline parallelism,'' in \emph{Advances in neural
  information processing systems}, 2019, pp. 103--112.

\bibitem{narayanan2019pipedream}
D.~Narayanan, A.~Harlap, A.~Phanishayee, V.~Seshadri, N.~R. Devanur, G.~R.
  Ganger, P.~B. Gibbons, and M.~Zaharia, ``Pipedream: generalized pipeline
  parallelism for dnn training,'' in \emph{Proceedings of the 27th ACM
  Symposium on Operating Systems Principles}, 2019, pp. 1--15.

\bibitem{harlap2018pipedream}
A.~Harlap, D.~Narayanan, A.~Phanishayee, V.~Seshadri, N.~Devanur, G.~Ganger,
  and P.~Gibbons, ``Pipedream: Fast and efficient pipeline parallel dnn
  training,'' \emph{arXiv preprint arXiv:1806.03377}, 2018.

\bibitem{raffel2019exploring}
C.~Raffel, N.~Shazeer, A.~Roberts, K.~Lee, S.~Narang, M.~Matena, Y.~Zhou,
  W.~Li, and P.~J. Liu, ``Exploring the limits of transfer learning with a
  unified text-to-text transformer,'' \emph{arXiv preprint arXiv:1910.10683},
  2019.

\bibitem{Jia2018Beyond}
Z.~Jia, M.~Zaharia, and A.~Aiken, ``Beyond data and model parallelism for deep
  neural networks,'' \emph{arXiv preprint arXiv:1807.05358}, 2018.

\bibitem{mirhoseini2017device}
A.~Mirhoseini, H.~Pham, Q.~V. Le, B.~Steiner, R.~Larsen, Y.~Zhou, N.~Kumar,
  M.~Norouzi, S.~Bengio, and J.~Dean, ``Device placement optimization with
  reinforcement learning,'' \emph{arXiv preprint arXiv:1706.04972}, 2017.

\bibitem{jax2018github}
\BIBentryALTinterwordspacing
J.~Bradbury, R.~Frostig, P.~Hawkins, M.~J. Johnson, C.~Leary, D.~Maclaurin, and
  S.~Wanderman-Milne, ``{JAX}: composable transformations of {P}ython+{N}um{P}y
  programs,'' 2018. [Online]. Available: \url{http://github.com/google/jax}
\BIBentrySTDinterwordspacing

\bibitem{Trax20}
T.~T. authors, \emph{Trax — Deep Learning with Clear Code and Speed}, 2020,
  \url{https://github.com/google/trax}.

\bibitem{xla19}
\emph{XLA: Optimizing Compiler for Machine Learning——Operation Semantics},
  2019, \url{https://www.tensorflow.org/xla/operation\_semantics}.

\bibitem{muchnick1997advanced}
S.~Muchnick \emph{et~al.}, \emph{Advanced compiler design
  implementation}.\hskip 1em plus 0.5em minus 0.4em\relax Morgan kaufmann,
  1997.

\bibitem{paszke2017automatic}
A.~Paszke, S.~Gross, S.~Chintala, G.~Chanan, E.~Yang, Z.~DeVito, Z.~Lin,
  A.~Desmaison, L.~Antiga, and A.~Lerer, ``Automatic differentiation in
  pytorch,'' 2017.

\bibitem{mnih2013playing}
V.~Mnih, K.~Kavukcuoglu, D.~Silver, A.~Graves, I.~Antonoglou, D.~Wierstra, and
  M.~Riedmiller, ``Playing atari with deep reinforcement learning,''
  \emph{arXiv preprint arXiv:1312.5602}, 2013.

\bibitem{bahdanau2014neural}
D.~Bahdanau, K.~Cho, and Y.~Bengio, ``Neural machine translation by jointly
  learning to align and translate,'' \emph{arXiv preprint arXiv:1409.0473},
  2014.

\bibitem{fan2020dapple}
S.~Fan, Y.~Rong, C.~Meng, Z.~Cao, S.~Wang, Z.~Zheng, C.~Wu, G.~Long, J.~Yang,
  L.~Xia \emph{et~al.}, ``Dapple: A pipelined data parallel approach for
  training large models,'' \emph{arXiv preprint arXiv:2007.01045}, 2020.

\bibitem{hessel2018rainbow}
M.~Hessel, J.~Modayil, H.~Van~Hasselt, T.~Schaul, G.~Ostrovski, W.~Dabney,
  D.~Horgan, B.~Piot, M.~Azar, and D.~Silver, ``Rainbow: Combining improvements
  in deep reinforcement learning,'' in \emph{Thirty-Second AAAI Conference on
  Artificial Intelligence}, 2018.

\bibitem{wang2016dueling}
Z.~Wang, T.~Schaul, M.~Hessel, H.~Hasselt, M.~Lanctot, and N.~Freitas,
  ``Dueling network architectures for deep reinforcement learning,'' in
  \emph{International conference on machine learning}, 2016, pp. 1995--2003.

\bibitem{fortunato2017noisy}
M.~Fortunato, M.~G. Azar, B.~Piot, J.~Menick, I.~Osband, A.~Graves, V.~Mnih,
  R.~Munos, D.~Hassabis, O.~Pietquin \emph{et~al.}, ``Noisy networks for
  exploration,'' \emph{arXiv preprint arXiv:1706.10295}, 2017.

\bibitem{chen2015mxnet}
T.~Chen, M.~Li, Y.~Li, M.~Lin, N.~Wang, M.~Wang, T.~Xiao, B.~Xu, C.~Zhang, and
  Z.~Zhang, ``Mxnet: A flexible and efficient machine learning library for
  heterogeneous distributed systems,'' \emph{arXiv preprint arXiv:1512.01274},
  2015.

\bibitem{DGX-1}
\emph{NVDIA DGX-1}, 2019,
  \url{https://www.nvidia.com/en-us/data-center/dgx-1/}.

\bibitem{nccl2019}
\emph{NCCL}, 2019, \url{https://developer.nvidia.com/nccl}.

\bibitem{devlin2018bert}
J.~Devlin, M.-W. Chang, K.~Lee, and K.~Toutanova, ``Bert: Pre-training of deep
  bidirectional transformers for language understanding,'' \emph{arXiv preprint
  arXiv:1810.04805}, 2018.

\bibitem{simonyan2014very}
K.~Simonyan and A.~Zisserman, ``Very deep convolutional networks for
  large-scale image recognition,'' \emph{arXiv preprint arXiv:1409.1556}, 2014.

\bibitem{kingma2014adam}
D.~P. Kingma and J.~Ba, ``Adam: A method for stochastic optimization,''
  \emph{arXiv preprint arXiv:1412.6980}, 2014.

\bibitem{schaul2015prioritized}
T.~Schaul, J.~Quan, I.~Antonoglou, and D.~Silver, ``Prioritized experience
  replay,'' \emph{arXiv preprint arXiv:1511.05952}, 2015.

\bibitem{van2016deep}
H.~Van~Hasselt, A.~Guez, and D.~Silver, ``Deep reinforcement learning with
  double q-learning,'' in \emph{Thirtieth AAAI conference on artificial
  intelligence}, 2016.

\bibitem{krizhevsky2014one}
A.~Krizhevsky, ``One weird trick for parallelizing convolutional neural
  networks,'' \emph{arXiv preprint arXiv:1404.5997}, 2014.

\bibitem{pal2019optimizing}
S.~Pal, E.~Ebrahimi, A.~Zulfiqar, Y.~Fu, V.~Zhang, S.~Migacz, D.~Nellans, and
  P.~Gupta, ``Optimizing multi-gpu parallelization strategies for deep learning
  training,'' \emph{IEEE Micro}, vol.~39, no.~5, pp. 91--101, 2019.

\bibitem{Cheng2017TensorFlow}
H.-T. Cheng, Z.~Haque, L.~Hong, M.~Ispir, C.~Mewald, I.~Polosukhin, G.~Roumpos,
  D.~Sculley, J.~Smith, D.~Soergel \emph{et~al.}, ``Tensorflow estimators:
  Managing simplicity vs. flexibility in high-level machine learning
  frameworks,'' in \emph{Proceedings of the 23rd ACM SIGKDD International
  Conference on Knowledge Discovery and Data Mining}, 2017, pp. 1763--1771.

\bibitem{zhan2019pipe}
J.~Zhan and J.~Zhang, ``Pipe-torch: Pipeline-based distributed deep learning in
  a gpu cluster with heterogeneous networking,'' in \emph{2019 Seventh
  International Conference on Advanced Cloud and Big Data (CBD)}.\hskip 1em
  plus 0.5em minus 0.4em\relax IEEE, 2019, pp. 55--60.

\bibitem{geng2019elasticpipe}
J.~Geng, D.~Li, and S.~Wang, ``Elasticpipe: An efficient and dynamic
  model-parallel solution to dnn training,'' in \emph{Proceedings of the 10th
  Workshop on Scientific Cloud Computing}, 2019, pp. 5--9.

\bibitem{yang2019pipemare}
B.~Yang, J.~Zhang, J.~Li, C.~R{\'e}, C.~R. Aberger, and C.~De~Sa, ``Pipemare:
  Asynchronous pipeline parallel dnn training,'' \emph{arXiv preprint
  arXiv:1910.05124}, 2019.

\bibitem{Goldie2020Placement}
A.~Goldie and A.~Mirhoseini, ``Placement optimization with deep reinforcement
  learning,'' in \emph{Proceedings of the 2020 International Symposium on
  Physical Design}, 2020, pp. 3--7.

\bibitem{mnih2015human}
V.~Mnih, K.~Kavukcuoglu, D.~Silver, A.~A. Rusu, J.~Veness, M.~G. Bellemare,
  A.~Graves, M.~Riedmiller, A.~K. Fidjeland, G.~Ostrovski \emph{et~al.},
  ``Human-level control through deep reinforcement learning,'' \emph{nature},
  vol. 518, no. 7540, pp. 529--533, 2015.

\end{thebibliography}


\begin{thebibliography}{10}
\providecommand{\url}[1]{#1}
\csname url@samestyle\endcsname
\providecommand{\newblock}{\relax}
\providecommand{\bibinfo}[2]{#2}
\providecommand{\BIBentrySTDinterwordspacing}{\spaceskip=0pt\relax}
\providecommand{\BIBentryALTinterwordstretchfactor}{4}
\providecommand{\BIBentryALTinterwordspacing}{\spaceskip=\fontdimen2\font plus
\BIBentryALTinterwordstretchfactor\fontdimen3\font minus
  \fontdimen4\font\relax}
\providecommand{\BIBforeignlanguage}[2]{{%
\expandafter\ifx\csname l@#1\endcsname\relax
\typeout{** WARNING: IEEEtran.bst: No hyphenation pattern has been}%
\typeout{** loaded for the language `#1'. Using the pattern for}%
\typeout{** the default language instead.}%
\else
\language=\csname l@#1\endcsname
\fi
#2}}
\providecommand{\BIBdecl}{\relax}
\BIBdecl

\bibitem{AI-and-compute}
Danny Hernandez. Dario Amodei, \emph{OpenAI-AI-and-compute}, 2019,
  \url{https://openai.com/blog/ai-and-compute/}.

\bibitem{rajbhandari2019zero}
S.~Rajbhandari, J.~Rasley, O.~Ruwase, and Y.~He, ``Zero: Memory optimization
  towards training a trillion parameter models,'' \emph{arXiv preprint
  arXiv:1910.02054}, 2019.

\bibitem{shoeybi2019megatron}
M.~Shoeybi, M.~Patwary, R.~Puri, P.~LeGresley, J.~Casper, and B.~Catanzaro,
  ``Megatron-lm: Training multi-billion parameter language models using gpu
  model parallelism,'' \emph{arXiv preprint arXiv:1909.08053}, 2019.

\bibitem{shazeer2018mesh}
N.~Shazeer, Y.~Cheng, N.~Parmar, D.~Tran, A.~Vaswani, P.~Koanantakool,
  P.~Hawkins, H.~Lee, M.~Hong, C.~Young, R.~Sepassi, and B.~Hechtman,
  ``{Mesh-TensorFlow}: Deep learning for supercomputers,'' in \emph{Neural
  Information Processing Systems}, 2018.

\bibitem{jia2018exploring}
Z.~Jia, S.~Lin, C.~R. Qi, and A.~Aiken, ``Exploring the hidden dimension in
  accelerating convolutional neural networks,'' 2018.

\bibitem{Geng2019HorizontalOV}
J.~Geng, D.~Li, and S.~Wang, ``Horizontal or vertical?: A hybrid approach to
  large-scale distributed machine learning,'' \emph{Proceedings of the 10th
  Workshop on Scientific Cloud Computing}, 2019.

\bibitem{dryden2019channel}
N.~Dryden, N.~Maruyama, T.~Moon, T.~Benson, M.~Snir, and B.~Van~Essen,
  ``Channel and filter parallelism for large-scale cnn training,'' in
  \emph{Proceedings of the International Conference for High Performance
  Computing, Networking, Storage and Analysis}, 2019, pp. 1--20.

\bibitem{huang2019gpipe}
Y.~Huang, Y.~Cheng, A.~Bapna, O.~Firat, D.~Chen, M.~Chen, H.~Lee, J.~Ngiam,
  Q.~V. Le, Y.~Wu \emph{et~al.}, ``Gpipe: Efficient training of giant neural
  networks using pipeline parallelism,'' in \emph{Advances in Neural
  Information Processing Systems}, 2019, pp. 103--112.

\bibitem{narayanan2019pipedream}
D.~Narayanan, A.~Harlap, A.~Phanishayee, V.~Seshadri, N.~R. Devanur, G.~R.
  Ganger, P.~B. Gibbons, and M.~Zaharia, ``Pipedream: generalized pipeline
  parallelism for dnn training,'' in \emph{Proceedings of the 27th ACM
  Symposium on Operating Systems Principles}, 2019, pp. 1--15.

\bibitem{abadi2016tensorflow}
M.~Abadi, P.~Barham, J.~Chen, Z.~Chen, A.~Davis, J.~Dean, M.~Devin,
  S.~Ghemawat, G.~Irving, M.~Isard \emph{et~al.}, ``Tensorflow: A system for
  large-scale machine learning,'' in \emph{12th $\{$USENIX$\}$ symposium on
  operating systems design and implementation ($\{$OSDI$\}$ 16)}, 2016, pp.
  265--283.

\bibitem{xla19}
T.~T. authors, \emph{XLA: Optimizing Compiler for Machine
  Learning——Operation Semantics}, 2019,
  \url{https://www.tensorflow.org/xla/operation\_semantics}.

\bibitem{muchnick1997advanced}
S.~Muchnick \emph{et~al.}, \emph{Advanced compiler design
  implementation}.\hskip 1em plus 0.5em minus 0.4em\relax Morgan kaufmann,
  1997.

\bibitem{buchlovsky2019tf}
P.~Buchlovsky, D.~Budden, D.~Grewe, C.~Jones, J.~Aslanides, F.~Besse, A.~Brock,
  A.~Clark, S.~G. Colmenarejo, A.~Pope \emph{et~al.}, ``Tf-replicator:
  Distributed machine learning for researchers,'' \emph{arXiv preprint
  arXiv:1902.00465}, 2019.

\bibitem{harlap2018pipedream}
A.~Harlap, D.~Narayanan, A.~Phanishayee, V.~Seshadri, N.~Devanur, G.~Ganger,
  and P.~Gibbons, ``Pipedream: Fast and efficient pipeline parallel dnn
  training,'' \emph{arXiv preprint arXiv:1806.03377}, 2018.

\bibitem{paszke2017automatic}
A.~Paszke, S.~Gross, S.~Chintala, G.~Chanan, E.~Yang, Z.~DeVito, Z.~Lin,
  A.~Desmaison, L.~Antiga, and A.~Lerer, ``Automatic differentiation in
  pytorch,'' 2017.

\bibitem{jax2018github}
\BIBentryALTinterwordspacing
J.~Bradbury, R.~Frostig, P.~Hawkins, M.~J. Johnson, C.~Leary, D.~Maclaurin, and
  S.~Wanderman-Milne, ``{JAX}: composable transformations of {P}ython+{N}um{P}y
  programs,'' 2018. [Online]. Available: \url{http://github.com/google/jax}
\BIBentrySTDinterwordspacing

\bibitem{Trax20}
T.~T. authors, \emph{Trax — Deep Learning with Clear Code and Speed}, 2020,
  \url{https://github.com/google/trax}.

\bibitem{mnih2013playing}
V.~Mnih, K.~Kavukcuoglu, D.~Silver, A.~Graves, I.~Antonoglou, D.~Wierstra, and
  M.~Riedmiller, ``Playing atari with deep reinforcement learning,''
  \emph{arXiv preprint arXiv:1312.5602}, 2013.

\bibitem{auer2009near}
P.~Auer, T.~Jaksch, and R.~Ortner, ``Near-optimal regret bounds for
  reinforcement learning,'' in \emph{Advances in neural information processing
  systems}, 2009, pp. 89--96.

\bibitem{bahdanau2014neural}
D.~Bahdanau, K.~Cho, and Y.~Bengio, ``Neural machine translation by jointly
  learning to align and translate,'' \emph{arXiv preprint arXiv:1409.0473},
  2014.

\bibitem{hessel2018rainbow}
M.~Hessel, J.~Modayil, H.~Van~Hasselt, T.~Schaul, G.~Ostrovski, W.~Dabney,
  D.~Horgan, B.~Piot, M.~Azar, and D.~Silver, ``Rainbow: Combining improvements
  in deep reinforcement learning,'' in \emph{Thirty-Second AAAI Conference on
  Artificial Intelligence}, 2018.

\bibitem{Wang2015Dueling}
Z.~Wang, T.~Schaul, M.~Hessel, H.~Van~Hasselt, M.~Lanctot, and N.~De~Freitas,
  ``Dueling network architectures for deep reinforcement learning,'' 2015.

\bibitem{fortunato2017noisy}
M.~Fortunato, M.~G. Azar, B.~Piot, J.~Menick, I.~Osband, A.~Graves, V.~Mnih,
  R.~Munos, D.~Hassabis, O.~Pietquin \emph{et~al.}, ``Noisy networks for
  exploration,'' \emph{arXiv preprint arXiv:1706.10295}, 2017.

\bibitem{chen2015mxnet}
T.~Chen, M.~Li, Y.~Li, M.~Lin, N.~Wang, M.~Wang, T.~Xiao, B.~Xu, C.~Zhang, and
  Z.~Zhang, ``Mxnet: A flexible and efficient machine learning library for
  heterogeneous distributed systems,'' \emph{arXiv preprint arXiv:1512.01274},
  2015.

\bibitem{DGX-1}
\emph{NVDIA DGX-1}, 2019,
  \url{https://www.nvidia.com/en-us/data-center/dgx-1/}.

\bibitem{nccl2019}
\emph{NCCL}, 2019, \url{https://developer.nvidia.com/nccl}.

\bibitem{devlin2018bert}
J.~Devlin, M.-W. Chang, K.~Lee, and K.~Toutanova, ``Bert: Pre-training of deep
  bidirectional transformers for language understanding,'' \emph{arXiv preprint
  arXiv:1810.04805}, 2018.

\bibitem{raffel2019Exploring}
C.~Raffel, N.~Shazeer, A.~Roberts, K.~Lee, S.~Narang, M.~Matena, Y.~Zhou,
  W.~Li, and P.~J. Liu, ``Exploring the limits of transfer learning with a
  unified text-to-text transformer,'' 2019.

\bibitem{simonyan2014very}
K.~Simonyan and A.~Zisserman, ``Very deep convolutional networks for
  large-scale image recognition,'' \emph{arXiv preprint arXiv:1409.1556}, 2014.

\bibitem{kingma2014adam}
D.~P. Kingma and J.~Ba, ``Adam: A method for stochastic optimization,''
  \emph{arXiv preprint arXiv:1412.6980}, 2014.

\bibitem{schaul2015prioritized}
T.~Schaul, J.~Quan, I.~Antonoglou, and D.~Silver, ``Prioritized experience
  replay,'' \emph{arXiv: Learning}, 2015.

\bibitem{van2016deep}
H.~Van~Hasselt, A.~Guez, and D.~Silver, ``Deep reinforcement learning with
  double q-learning,'' in \emph{Thirtieth AAAI conference on artificial
  intelligence}, 2016.

\bibitem{krizhevsky2014one}
A.~Krizhevsky, ``One weird trick for parallelizing convolutional neural
  networks,'' \emph{arXiv preprint arXiv:1404.5997}, 2014.

\bibitem{pal2019optimizing}
S.~Pal, E.~Ebrahimi, A.~Zulfiqar, Y.~Fu, V.~Zhang, S.~Migacz, D.~Nellans, and
  P.~Gupta, ``Optimizing multi-gpu parallelization strategies for deep learning
  training,'' \emph{IEEE Micro}, vol.~39, no.~5, pp. 91--101, 2019.

\bibitem{Cheng2017TensorFlow}
H.~T. Cheng, D.~Soergel, T.~Yuan, P.~Tucker, and J.~Smith, ``Tensorflow
  estimators: Managing simplicity vs. flexibility in high-level machine
  learning frameworks,'' in \emph{Acm Sigkdd International Conference}, 2017.

\bibitem{Jia2018Beyond}
Z.~Jia, M.~Zaharia, and A.~Aiken, ``Beyond data and model parallelism for deep
  neural networks,'' 2018.

\bibitem{zhan2019pipe}
J.~Zhan and J.~Zhang, ``Pipe-torch: Pipeline-based distributed deep learning in
  a gpu cluster with heterogeneous networking,'' in \emph{2019 Seventh
  International Conference on Advanced Cloud and Big Data (CBD)}.\hskip 1em
  plus 0.5em minus 0.4em\relax IEEE, 2019, pp. 55--60.

\bibitem{geng2019elasticpipe}
J.~Geng, D.~Li, and S.~Wang, ``Elasticpipe: An efficient and dynamic
  model-parallel solution to dnn training,'' in \emph{Proceedings of the 10th
  Workshop on Scientific Cloud Computing}.\hskip 1em plus 0.5em minus
  0.4em\relax ACM, 2019, pp. 5--9.

\bibitem{yang2019pipemare}
B.~Yang, J.~Zhang, J.~Li, C.~Ré, C.~R. Aberger, and C.~De~Sa, ``Pipemare:
  Asynchronous pipeline parallel dnn training,'' 2019.

\bibitem{Goldie2020Placement}
A.~Goldie and A.~Mirhoseini, ``Placement optimization with deep reinforcement
  learning,'' 2020.

\bibitem{mnih2015human}
V.~Mnih, K.~Kavukcuoglu, D.~Silver, A.~A. Rusu, J.~Veness, M.~G. Bellemare,
  A.~Graves, M.~Riedmiller, A.~K. Fidjeland, G.~Ostrovski \emph{et~al.},
  ``Human-level control through deep reinforcement learning,'' \emph{nature},
  vol. 518, no. 7540, pp. 529--533, 2015.

\bibitem{wang2016dueling}
Z.~Wang, T.~Schaul, M.~Hessel, H.~Hasselt, M.~Lanctot, and N.~Freitas,
  ``Dueling network architectures for deep reinforcement learning,'' in
  \emph{International conference on machine learning}, 2016, pp. 1995--2003.
  
\end{thebibliography}

\end{document}